\documentclass[12pt]{article}
\usepackage[utf8]{inputenc}
\usepackage[authoryear]{natbib}
\usepackage{fullpage}
\usepackage{graphicx}
\usepackage{todonotes} 
\usepackage{dcolumn}
\usepackage{amsmath}
\usepackage{hyperref} 
\usepackage[capitalize]{cleveref}
\usepackage{caption}
\usepackage{subcaption}
\usepackage{longtable}

\usepackage{setspace}
\usepackage{xcolor}
\newcommand{\edit}[1]{\textcolor{black}{#1}}
\newcommand{\moved}[1]{\textcolor{black}{#1}}
\usepackage{chngcntr}

\usepackage{amsfonts,amssymb}\usepackage{booktabs}
\newcounter{todocounter}

  \usepackage{makecell}

\usepackage{titling}
\usepackage{setspace}
\usepackage{authblk}
\def\spacingset#1{\renewcommand{\baselinestretch}%
{#1}\small\normalsize} \spacingset{1}
\spacingset{1}

\title{Synthetic Control Analysis of the Short-Term Impact of New York State's Bail Elimination Act on Aggregate Crime 
}
\author[1]{Angela Zhou\footnote{Authors are listed nearly alphabetically by first name. }\thanks{\texttt{zhoua@usc.edu}}}
\author[2]{Andrew Koo\thanks{\texttt{alk272@cornell.edu}}}
\author[2]{Nathan Kallus\thanks{\texttt{kallus@cornell.edu}}}
\author[3]{Rene Ropac\thanks{\texttt{rropac@nycja.org}, at CJA when most of the paper was completed. }}
\author[3]{Richard Peterson\thanks{\texttt{RPeterson@nycja.org}, at CJA when most of the paper was completed. }}
\author[3]{Stephen Koppel\thanks{\texttt{SKoppel@nycja.org}, at CJA when most of the paper was completed. }}
\author[3]{Tiffany Bergin\thanks{\texttt{tbergin@nycja.org}}, at CJA when most of the paper was completed. }
\affil[1]{University of Southern California}
\affil[2]{Cornell University/Cornell Tech}
\affil[3]{New York City Criminal Justice Agency}

\begin{document}

\maketitle

\begin{abstract}
    We conduct an empirical evaluation of the short-term impact of New York's bail reform on crime. New York State's Bail Elimination Act went into effect on January 1, 2020, eliminating money bail and pretrial detention for nearly all misdemeanor and nonviolent felony defendants. Our analysis of effects on aggregate crime rates after the reform informs the understanding of bail reform and general deterrence, rather than specific deterrence via re-arrest rates of the detained/released population. We conduct a synthetic control analysis for a comparative case study of the impact of bail reform. We focus on synthetic control analysis of post-intervention changes in crime for assault, theft, burglary, robbery, and drug crimes, constructing a dataset from publicly reported crime data of 27 large municipalities. Due to the short time frame before the onset of COVID-19 and its far-reaching effects, we restrict attention to a short post-intervention time period. Nonetheless, evaluation of short-term impacts may still inform hypotheses of general deterrence of bail reform policy. Our findings, including placebo checks and other robustness checks, show that for assault, theft, and drug crimes, there is no significant impact of bail reform on aggregate crime. For robbery, we find a statistically significant increase; for burglary, the synthetic control is more variable and our analysis is deemed less conclusive. Since our study assesses the short-term impacts, further work studying long-term impacts of bail reform and on specific deterrence remains necessary. 
\end{abstract}

\section{Introduction}\label{sec:intro}

The use of cash bail has come under increasing scrutiny in the United States, with several states and jurisdictions adopting or considering reforms to limit its use \citep{covert2017nation}. One key question is whether such reforms -- which generally lead to fewer pretrial detainees -- impact crime rates. We address this question by investigating a real-life case study, New York's implementation of a sweeping bail reform legislation which took effect on January 1, 2020.
We focus on the impact on the incidence of crime of different types in New York City (NYC), using the synthetic control method \citep{abadie2010synthetic,abadie2011synth} to conduct comparative event-study analysis. 

Specifically, our analysis takes advantage of the fact that cities across the United States historically experience similar shifts in crime trends 
\citep{b10}%
, which allows us to compare the developments in NYC to a pool of control jurisdictions where no such policy changes were made within the timeframe of our study between January 1, 2018 and March 15, 2020. Our focus is specifically on the effect of New York State's Bail Elimination Act on New York City crime rates. ``Bail reform" is a broad term that is applied to a wide range of policies, from nudges to judicial actors to reduce the use of money bail, to a wide-ranging bill such as the Bail Elimination Act which eliminate the use of money bail for a wide variety of charges without judicial discretion. %
Our conceptualization of treatment is the specific elimination of money bail in New York State as mandated by the Bail Elimination Act.

This cross-unit comparison aids causal inference by disentangling the true impact of bail reform from other concurrent events and trends, which would not be possible by conducting a within-unit comparison such as interrupted-time series 
\citep{degli2020can},
which we also conduct as a reference starting point.
To facilitate our analysis, we collected data directly from 27 different cities' police departments on the incidence rates of different types of crimes. Using the synthetic control method, for each crime type, we construct a weighted combination of these cities to match NYC before 2020. Roughly, comparing the divergence -- or lack thereof -- in 2020 allows us to assess the impact of the novel treatment in 2020.
Our analysis is limited to short-term impacts due to the onset of COVID-19 shortly after the institution of bail reform. In \Cref{sec:limitations,sec:conclusion} we discuss limitations in further detail. Overall, following statistical inference and robustness checks, we find little to no evidence of any significant impact of bail reform on total crime rates. The synthetic control explains away observed trends in post-intervention crime rates, assessing the average treatment effect estimate against the variability in the distribution of effect estimates under placebo analyses with known null interventional effect. These results are robust under other robustness checks. %

The structure of our paper is as follows: we first discuss the background, data, methods, and results, before summarizing limitations of the analysis and offering concluding remarks. We review the context of bail reform and of findings elsewhere as well as the specific changes under the reform passed in New York, and we highlight domain-level considerations that inform methodological choices for our empirical evaluation and our focus on a synthetic control evaluation. We then turn to describing data, methods, and our results: we describe our constructed dataset and pre-processing as well as our main analyses. We focus on synthetic control analysis of post-intervention changes in crime for assault, theft, burglary, robbery, and drug crimes. Our findings, including placebo checks and other robustness checks, show that for assault, theft, and drug crimes, there is no significant impact of bail reform on crime; for burglary and robbery, we similarly have null findings but the synthetic control is also more variable so these are deemed less conclusive. Finally, we discuss important limitations of our study in the operational context of New York before concluding by presenting directions for future research.

\section{Background}\label{sec:background}

\subsection{Bail reform}
As more U.S. jurisdictions consider reforms to end or sharply reduce the use of cash bail, debates about the impacts of such reforms have gained urgency. Key questions include the effects of such reforms on: (i) pretrial detention rates; (ii) pretrial rearrest rates; and (iii) overall crime rates. The first two questions have received the most attention in the literature. This study makes a unique contribution by focusing on the third question. Specifically, in assessing the effects on aggregate crime rates, we explore the broader consequences of bail reform, including potential consequences for general deterrence.

To assess bail reform’s impacts on crime and rearrest, we investigate the effects of New York State’s Bail Elimination Act, which went into effect on January 1, 2020. Other reports with more descriptive detail and statistics include those of \cite{rempel2020bail} and \cite{rempel2020oneyearlater}. The reform ``eliminates money bail and pretrial detention for nearly all misdemeanor and nonviolent felony defendants," among other policy changes, such as directing judges to consider ability to pay when setting money bail amounts \citep[p. 1]{rempel2019bail}. Prior to its enactment, judges in New York had the option to set bail or remand in all cases—regardless of the charge a defendant was facing. The new law proscribes money bail in almost all cases charged with a misdemeanor (except for sex offenses and domestic violence, which make up about 12\% of those charged with misdemeanors) and disallows remand for misdemeanors \citep[p.  2]{rempel2019bail}. For felonies, in summary, the new criteria where both money bail and remand are possible ``permit bail and detention with nearly all violent felonies but rule it out with nearly all nonviolent felonies" \citep[p. 2]{rempel2019bail}. In effect, this meant that pretrial release was mandated in nearly all misdemeanor and nonviolent felony cases on the basis of a single factor: the present charge in a case.

We highlight some descriptive statistics which shed light on magnitudes of differences in charge eligibility under the reform. In 2019, ``collectively, judges ordered bail or remand in 7 percent of misdemeanors, 35 percent of nonviolent felonies, and 62 percent of violent felonies." \citep[p. 9]{rempel2019bail}. Of the cases made bail-eligible, judges ordered bail or remand in 54\% of the cases, and of the cases made ineligible for bail, judges used bail or remand 12\% of the time. Where judges ordered bail or remand in 2019, in about 13,000 cases they would be unable to do so under the original reform; the greatest difference is in non-violent felonies \citep[p. 10]{rempel2019bail}.

Partly due to concerns related to public safety, six months after the law went into effect, on July 2, 2020, it was amended to expand the eligibility criteria \citep{nybailreform,bailreformamendments}. The amendments included a list of newly eligible misdemeanor and non-violent felony charges, as well as broader eligibility criteria related to a defendant’s criminal history, including whether a defendant had a prior history of felony convictions, a separate pending case, or was on probation or parole. These changes effectively ended New York’s charge-based approach to bail reform, replacing it with a second bail reform regime in which bail eligibility was determined based on the present charge or a defendant’s criminal history.

\subsection{Evaluations of bail reform}\label{sec:apx-bailreformcontext}
\moved{
The earliest and best known example of bail reform in the U.S. is Washington, D.C., which in 1992 replaced cash bail with a risk-based system \citep{lockwood2020reform}. Although no rigorous evaluations were conducted at the time, more recent descriptive statistics reveal a pretrial rearrest rate of 12-14\%, which is broadly in line with that of other jurisdictions that still use cash bail \citep{appr20}.
 In 2017, New Jersey implemented a raft of pretrial reforms, including: (1) mandating the use of a pretrial risk assessment tool, (2) restricting the use of bail to defendants at high risk of flight, and (3) restricting pretrial detention to defendants at high risk of either flight or new criminal activity. While no rigorous studies have yet been conducted on the impact of the reforms, descriptive analyses suggest
 that after the reform went into effect pretrial rearrest rates did not subsequently rise \citep{njcourts2020}.}
\moved{
In 2017, Philadelphia's District Attorney announced that his office would no longer seek bail for 25 low-level felony and misdemeanor offenses. Using an instrumental difference-in-differences design to compare bail eligible offenses to bail ineligible offenses, researchers found that the policy had no impact on the likelihood of either pretrial detention or pretrial rearrest \citep{ouss2020bail}. Also in 2017, the Chief Justice in Cook County, Illinois, issued a court order (G.0. 18.8A), establishing a presumption of release for most defendants. Using multivariate models to compare matched samples of defendants pre- and post-reform, researchers found that the order was associated with a 4 percentage-point decrease in pretrial detention but no change in pretrial rearrest \citep{sa20}. Similar findings were described in an official report \citep{judge2019}, but another study of the same reform found that crime among individuals on pretrial release did indeed increase \citep{cassell2020bail}.
In 2019, in Harris County, Texas, a court-ordered consent decree (Rule 9) mandated the pretrial release of a subset of misdemeanor offenses. A descriptive analysis of 8 pretrial rearrest rates among misdemeanor defendants showed no change following the implementation of the order \citep{harriscounty2020}. In 2014, Mecklenburg, North Carolina, a new automated risk assessment tool was rolled out to replace an older risk assessment instrument. Using an interrupted time series design, researchers found that the implementation was followed by an 11 percentage-point decrease in money bail, and a 2 percentage-point increase in pretrial rearrest \citep{mecklenburg2019reform}.}

Outside of any reform context, research has also found that bail and pretrial detention are associated with a higher likelihood of rearrest or recidivism \citep{monaghan2020bail,gupta2016bail,lowenkamp2013detention}. Other research has highlighted pretrial detention’s harmful collateral impacts, including its impact on future employment and residential stability for arrested individuals \citep{dobbie2018detention,heaton2017detention}. Additionally, bail and pretrial detention can also intensify racial disparities at later phases of the criminal legal process \citep{donnelly2018bail}.

New York's bail reform is a difference in kind, not only in degree, from other reforms. 
In the typology of \cite[p. 5-7]{jorgensen2021current}'s landscape analysis of bail reforms, in categorizing state-led reforms, during our study period, New York stands alone in the category of ``abolish[ing] cash bail for some or all crimes". (Other states include Illinois and Maine, whose reforms go into effect strictly after our study period). Among city- and county-led reforms, most such reforms are outside our study period. We discuss in greater detail in \Cref{sec-apx-addldiscussion-bailreforms}.

Finally, we clarify how our study is relevant to general deterrence by studying aggregate crime. There are two different mechanisms to conceptualize impacts of bail reform on crime: \textit{general deterrence}, which changes the costs/benefits of arrest for committing crime and hence possibly inducing increases in crime rates \edit{among the general population}, and \textit{specific deterrence}, regarding incapacitative impacts of pretrial-detention and bail policy on individuals or \edit{reducing pretrial re-arrest and/or recidivism specifically in the pretrial population}. Our analysis does not inform specific deterrence.%

On the one hand, increasing the released population may generally be expected to increase the absolute magnitude of re-arrests, since bail (via detention) has incapacitative effects. Setting monetary bail may also pose financial incentives to prevent reoffense, leading to specific deterrence. The recent analysis of \citet{albright2021no} studies the question of financial incentives in a different context. 
Therefore, bail reform could introduce societal costs. If bail reform leads to an increase in crime via specific deterrence or via general deterrence, assessing the absolute magnitude helps articulate tradeoffs against other important arguments for bail reform. \citet{dobbie2021us} overviews these considerations balancing individual rights and public interests. On the other hand, perhaps bail reform does not increase aggregate crime (does not reduce general deterrence), so that there are fewer than posited societal costs of bail reform weighed against individual-level benefits. \edit{Our analysis indeed suggests this later case. Such benefits, not included in our quantitative analysis,} include avoiding the distributional consequences on detention for those unable to make money bail, civil liberties, racial and ethnic disparities, constitutional and moral arguments, and other important policy considerations.

This study represents the first attempt to evaluate the effects of the bail reform of New York State's Bail Elimination Act on aggregate crime rates. Bail reform is difficult to evaluate due to (i) confounders (other 
changes that take place at the same time leading to cross-unit unobserved confounding); (ii) data limitations and difficulties of model specification; and (iii) the uniqueness of particular case studies (findings that hold true in NYC may not hold true elsewhere and therefore are not representative of bail reform's generalized effects). \edit{We address (i) and (iii) by using synthetic control methodology \citep{abadie2010synthetic}, which has been used to evaluate policy, including on crime \citep{saunders2015synthetic,donohue2019right,rees2019little,ben2021effect}. We address (ii) by aggregating open data from municipal police departments. Synthetic control can be argued to estimate causal effects under shared unobserved confounding, while interrupted time series cannot. Unobserved confounders could include broader, shared shifts in crime incidence  such as federal or county-level policy and aforementioned mechanisms discussed in the previous section driving crime rates. In \Cref{sec:apx-addldiscussion}, we include a discussion on the criminological theory of crime trends as a national phenomenon.} The synthetic control reweights the non-NYC municipal police departments in order to track NYC's pre-intervention trajectory. Divergence in outcomes between NYC and the synthetic control (weighted composite of other cities) allows for a comparative analysis that uses the synthetic control to account for shared aggregate-level changes in crime rates. We include more detail on dataset construction in \Cref{sec:data} and the synthetic control method in \Cref{sec:methods}.

\section{Overview of  data}\label{sec:data}

We construct a dataset by compiling crime data released by police departments across 27 cities across the United States from the period January 1, 2018 - March 15, 2020 for the purpose of establishing a synthetic control for NYC crime trends. Although the main analysis is synthetic control, we also conduct an interrupted time series analysis with only New York City's data starting a year prior, from January 1, 2017, as we are less restricted by ensuring data availability across all municipalities for interrupted time series.

Reported incidents of crime were obtained from police department-reported incidents accessible via data portals of the 30 most populous U.S. cities (table of sources in \Cref{sec:apx-datapreprocessing}). \edit{We omitted Dallas, Atlanta and Fort Worth due to significant reporting discontinuities in the data. We removed Philadelphia, San Francisco, and Houston due to significant bail reform that occurred during the study period. We also removed some cities due to poor pre-treatment fit as suggested in \citep{abadie2010synthetic}. Other locations passed bail reforms before, or very near, the start of our synthetic control timeframe on January 1, 2018. These locations are retained in our analysis under an assumption that their prior treatment effects of local reforms were realized by the beginning of our study period. See \Cref{sec:apx-datapreprocessing} for more details.}

Our use of such incident report data comes with standard disclaimers regarding crime data and potential measurement issues: see \citet{ucrbook} for extensive discussion of particularities with this type of data. For our comparative case study, measurement biases that do not differ across units do not affect analysis conclusions. 

Each row in this dataset corresponds to an incident of crime. The dataset was constructed by merging the reported incidents from each agency. However, each agency reports crime types in a slightly different way, via different descriptors and categorizations. \edit{Our approach based on aggregating temporally fine data (weekly) differs from standard approaches that use FBI Uniform Crime Reporting (UCR) data at a monthly timescale.} In our dataset, some agencies, but not all, report UCR codes, but at different granularities. We therefore code crime types and construct a schema that broadly resembles UCR's index crime designations. The crimes at the second level of the hierarchy, at which level we conducted our analysis are: 
\edit{
\begin{itemize}
    \item Homicide: Murder and manslaughter (does not include justifiable homicide)
    \item Robbery: All taking of property through force or threat of force
    \item Assault: Includes aggravated and simple assaults when possible
    \item Burglary: Unlawful entry into structure (residential and commercial) to commit theft without use of force
    \item Theft: Unlawful taking of property (includes motor vehicle theft)
    \item Drug: All drug-related abuses including cultivation, distribution, sale, purchase, use, possession, transportation, or importation of any controlled drug or narcotic substance.
\end{itemize}
}
\edit{
\begin{table}[]
\caption{Summary of crime classifications}\label{table-crime-classifications}
\begin{tabular}{@{}lllll@{}}
\toprule
Assault                                                                                    & Theft                                                                                                                     & Burglary                                                                                               & Drug & Robbery  \\ \midrule
\begin{tabular}[c]{@{}l@{}}Aggravated Assault\\ Simple Assault\\ Unclassified\end{tabular} & \begin{tabular}[c]{@{}l@{}}Grand Larceny Auto\\ Grand Larceny\\ Theft from Auto\\ Other Theft\\ Unclassified\end{tabular} & \begin{tabular}[c]{@{}l@{}}Residential Burglary\\ Non-Residential Burglary\\ Unclassified\end{tabular} &      &               \\ \bottomrule
\end{tabular}
\end{table}
}
These definitions are modified from other crime frameworks, such as a previous Bureau of Justice Statistics analysis\footnote{\url{https://www.bjs.gov/recidivism/}}, FBI offense definitions \footnote{\url{https://ucr.fbi.gov/crime-in-the-u.s/2019/crime-in-the-u.s.-2019/topic-pages/offense-definitions}}, and definitions from City Crime Stats \footnote{\url{https://papers.ssrn.com/sol3/papers.cfm?abstract_id=3674032}}. \edit{See \Cref{table-crime-classifications} for a summary of what these categories include.} This study was conducted before the 2020 UCR data were available. Moreover, UCR data reporting is at the monthly level, which would result in only three post-intervention timepoints at the timescale of our post-intervention, pre-COVID window, muddying any inference. In contrast, constructing the dataset at a daily reporting level allows for greater flexibility and improves the precision of our estimates and inferences. %
In \Cref{sec:apx-datasetdescription-subUCR} we provide a more granular comparison.

Bail reform went into effect on January 1, 2020, which we use as the intervention date. However, the post-intervention period is necessarily abbreviated due to COVID lockdowns and COVID-related changes which introduce a large degree of nonstationarity. \edit{The pandemic was accompanied by several developments that could be plausibly linked to crime, such as increased economic hardship \citep{pcmww20}%
, decarceration \citep{p21}%
, an uptick in gun purchases \citep{smpstablw21}%
 and gun carrying \citep{aa21}%
, and halted non-policing efforts to prevent violence \citep{p20,bb20}. Including post-pandemic data in our analysis could therefore introduce treatment effects from the pandemic and differential responses to it across municipalities.} In NYC, the lockdown for COVID went into effect on March 15, 2020 which we use as the end of the post-intervention period. A partial rollback of bail reform rollback went into effect in New York on July 2, 2020, altogether outside our post- or pre-treatment windows.

Our treatment of interest is specifically the effect of \textit{New York State's bail reform} on New York City, which eliminated money bail and pretrial detention for nearly all misdemeanor and nonviolent felony defendants. We only include other units whose other bail reform packages occurred prior to the synthetic control timeframe, in \Cref{sec-apx-addldiscussion-bailreforms} we discuss more precise differences.

\section{Methods}\label{sec:methods}

Our preferred methodology is synthetic control: we now describe its benefits in this section as compared to other event-study methods. Relative to comparative case study approaches that would require choosing a comparator municipality, using synthetic control allows for flexible construction of a comparison, based on fitting the trajectory of crime rates before the intervention. This is helpful because the cities themselves differ on a number of factors (seasonality patterns, opportunities for different types of crime, etc.) so that manually choosing a control comparator is not self-evident in this setting. 

Relative to a within-NYC time series analysis, synthetic control allows for accounting for possible co-movements in underlying crime trends (for example, changes in federal policy or broader economic trends) with other non-treated units; the possibility of such trends is discussed in \Cref{sec:background}. Further, relative to interrupted time series, comparative interrupted time series, or differences-in-differences analysis, an important consideration in favor of synthetic control is that it compares the post-treatment time period to a longer pre-treatment time period than would occur otherwise, under segmented regression used for the previous methods. At a domain-level, the crucial assumption of \textit{no-anticipation} is violated. In our setting, no-anticipation presumes that outcomes do not change due to anticipation of the actual intervention date. However, anticipation was already quantitatively documented by \citet{cci_summary} with increased rates of supervised release in the weeks and months preceding the January 1 implementation of the reform; judges avoided processing individuals for bail or detention who would have been released under the new regime soon afterwards. \edit{To conclude, synthetic control is a more appropriate methodology that will additionally handle 1) shared unobserved confounders (such as documented crime trends) and 2) documented violations of a \textit{no-anticipation} assumption and is preferred to a comparative case study.} %

The original synthetic control method \citep{abadie2010synthetic} uses outcome predictors in order to balance a weighted distance between predictors of treated unit and predictors of control units. We first introduce the terminology and notation of the synthetic control, before describing the variants we introduce following the literature. 

Index the units (here, municipalities) as $j=1,2,\dots,J+1$ and associate the first index, $j=1$, to the treated unit (NYC). The possible ``donor pool" is a collection of untreated units, $j=2,\dots,J+1$, not affected by the intervention (bail reform). The data comprises $T$ time periods, with $T_0$ of these periods occurring before the intervention.
For each unit and time period, an outcome $Y_{jt}$ is measured (aggregate crime level within a categroy).
Let $Y_{jt}(1)$ denote the potential response under the treatment (bail reform intervention) and $Y_{jt}(0)$ otherwise, that is, under control. The effect of the intervention of interest for the treated unit at some time $t> T_0$ is $$\tau_{1t} = Y_{1t}(1) -Y_{1t}(0).$$ Of course, the fundamental problem of causal inference is that for any time $t>T_0$ we only observe $Y_{1t}(1)$, not the post-intervention outcome without the intervention. 

The synthetic control method constructs an estimate for $Y_{1t}(0)$ as a reweighted average of donor units' outcomes, $Y_{jt}(0)$, for whom we \textit{do} observe outcomes under non-intervention after the intervention time. The synthetic control estimate is specified via a \textit{weight vector} $W=[w_2, \dots, w_{J+1}]$ that is used to construct a weighted average of donor units (the synthetic control). 
We optimize for the synthetic control weights minimizing the predicted mean-square error for pre-intervention outcomes $Y$ plus a ridge regularization: 

\begin{equation}
    w^* \in \arg\min \left\{ \frac{1}{T_0}\sum_{t\leq T_0}  \left(Y_{1t} - \sum_{j=2}^{J+1} w_j Y_{jt}\right)^2  + \lambda \| w\|_2  \colon \sum_{j=2}^{J+1} w_j = 1 \right\} 
\end{equation}
The above program is easily solvable as a convex quadratic program. Then the synthetic control \textit{estimates} of the counterfactual and treatment effect are, respectively: 
$$ \hat{Y}_{1t}(0) = \sum_{j=2}^{J+1} w^*_j Y_{jt}, \qquad \hat\tau_{1t} = Y_{1t}(1) - \hat Y_{1t}(0)  $$

Our specification differs somewhat from \citet{abadie2011synth} and but follows common choices used in the literature. 
We implement synthetic control with unlagged outcomes and without covariates. (Including covariates would be either too noisy or redundant \citep{kaul2015synthetic}.)
For better interpretability and to avoid hyperparameter tuning,
we do not consider re-weighting mean-squared errors as in the original paper, i.e., a reweighting hyperparameter across donor units. We include a ridge regression penalty, following a suggestion of \citet{abadie2015comparative},
and allow negative weights. \citet{doudchenko2016balancing} discusses some of these relaxations. %

Due to the widely varying absolute magnitudes of crime in different cities --- since NYC is the most populated municipality --- we normalize the crime series in each city by the city's population (not metropolitan statistical area). Similarly, \citet{abadie2010synthetic} considers \textit{per-capita} smoking rates. We also demean the trends in order to build the synthetic control weights, subtracting the pre-treatment average for each city. This normalization is crucial for our permutation-based placebo checks down the line, as we assess the magnitude of the treatment effect for NYC relative to the magnitude of the treatment effect assessed under placebo checks (more details in \Cref{sec:results}).

\begin{table}[!t] \centering 
  \caption{Data descriptives: crime time series pre-intervention, daily counts}
  \label{table:its-descriptives-pre} 
\begin{tabular}{@{\extracolsep{5pt}}lcccccc} 
\\[-1.8ex]\hline 
\hline \\[-1.8ex] 
Statistic & \multicolumn{1}{c}{Mean} & \multicolumn{1}{c}{St. Dev.} & \multicolumn{1}{c}{Min} & \multicolumn{1}{c}{Pctl(25)} & \multicolumn{1}{c}{Pctl(75)} & \multicolumn{1}{c}{Max} \\ 
\hline \\[-1.8ex] 
Assault  & 246.566 & 29.849 & 135 & 226 & 267 & 359 \\ 
Theft  & 373.182 & 50.718 & 150 & 339 & 410 & 517 \\ 
Burglary  & 31.537 & 7.571 & 12 & 26 & 37 & 57 \\ 
Drug  & 45.642 & 21.013 & 1 & 30 & 60 & 105 \\ 
Robbery  & 36.747 & 8.082 & 18 & 31 & 42 & 71 \\ 
Homicide  & 0.841 & 1.054 & 0 & 0 & 1 & 9 \\ 
\hline \\[-1.8ex] 

\end{tabular} 

\end{table} 

\begin{table}[!t] \centering 
  \caption{Data descriptives: crime time series post-intervention, daily counts}
  \label{table:its-descriptives} 
\begin{tabular}{@{\extracolsep{5pt}}lcccccc} 
\\[-1.8ex]\hline 
\hline \\[-1.8ex] 
Statistic &  \multicolumn{1}{c}{Mean} & \multicolumn{1}{c}{St. Dev.} & \multicolumn{1}{c}{Min} & \multicolumn{1}{c}{Pctl(25)} & \multicolumn{1}{c}{Pctl(75)} & \multicolumn{1}{c}{Max} \\ 
\hline \\[-1.8ex] 
Assault  & 250.068 & 21.694 & 197 & 234.2 & 265 & 318 \\ 
Theft  & 381.851 & 35.118 & 302 & 358.2 & 404.5 & 462 \\ 
Burglary  & 36.135 & 7.206 & 19 & 31 & 41.5 & 52 \\ 
Drug  & 39.149 & 17.633 & 8 & 22.2 & 51.8 & 73 \\ 
Robbery  & 40.351 & 7.023 & 19 & 36 & 44 & 58 \\ 
Homicide  & 0.797 & 1.033 & 0 & 0 & 1 & 4 \\ 
\hline \\[-1.8ex] 
\end{tabular} 
\end{table}

\section{Results}\label{sec:results} 

Our primary analysis is with synthetic control. We assess uncertainty via placebo checks and other robustness checks that are common in the literature. In \Cref{sec:apx-its}, for completeness we also include an interrupted time series analysis.

\subsection{Implementation and details}\label{sec:results-synth-control}
We first describe data pre-processing steps used to ensure quality of the synthetic control fit, as well as describe our use of placebo checks common in the synthetic control literature to quantify the uncertainty in the treatment effect estimate, before describing the inferential results.

We consider estimates of an average treatment effect by averaging post-intervention outcomes to estimate the treated outcome and estimating the control outcome for NYC by reweighting with the synthetic control weights:  
\begin{equation}\label{eqn:ate}
    \hat\tau = \frac{1}{T-T_0}
    \left(\sum_{t=T_0}^{T} Y_{1t} - \sum_{j=1}^{J+1} w_j^* Y_{jt}
    \right) 
\end{equation}

We analyze Assault, Theft, Drug, Robbery, and Burglary crimes. Assault and Robbery are categorized as violent crimes while Burglary and Theft are categorized as property crimes. \edit{The number of homicides is quite small so that the analysis for synthetic control is too noisy on this timescale. For example, \Cref{table:its-descriptives-pre,table:its-descriptives} show it is three order of magnitudes more sparse than burglary and robbery. Due to data sparsity and noisy outcomes, we aggregate different crime types to different temporal aggregations by summing weekly counts over longer time intervals. Assault and Theft crimes remain on a weekly aggregation, Burglary and Robbery crimes are aggregated to 5 weeks, while Robbery crimes are aggregated to 2 weeks and Drug crimes to 3 weeks.}

The data-pre-processing steps are fairly standard in the synthetic control literature, and include tuning the ridge penalty by pre-treatment prediction quality (RMSE, \edit{root mean-squared error}), determining aggregation level for outcome series by pre-treatment fit, omitting cities based on pre-treatment fit more than 7.5 times that of the original synthetic control (as suggested in \citep{abadie2011synth}). In the interest of brevity, we  discuss data-driven or other justifications in \Cref{sec:apx-datapreprocessing}.

\subsection{Results: Linear regression with ridge penalty specification}

\begin{figure}[!htp]
\begin{subfigure}[t]{0.5\linewidth}
\includegraphics[width=\textwidth]{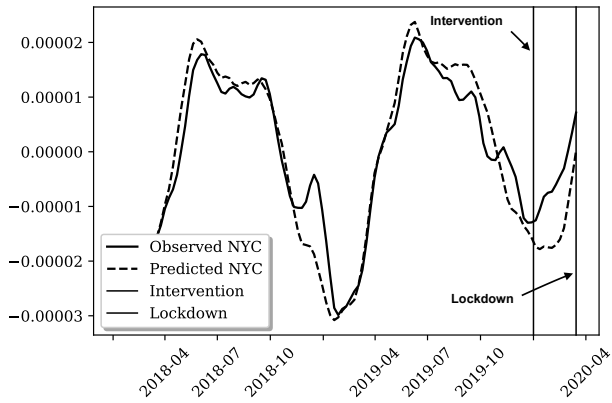}\caption{Violent Assault}\end{subfigure}\begin{subfigure}[t]{0.5\linewidth}\includegraphics[width=\textwidth]{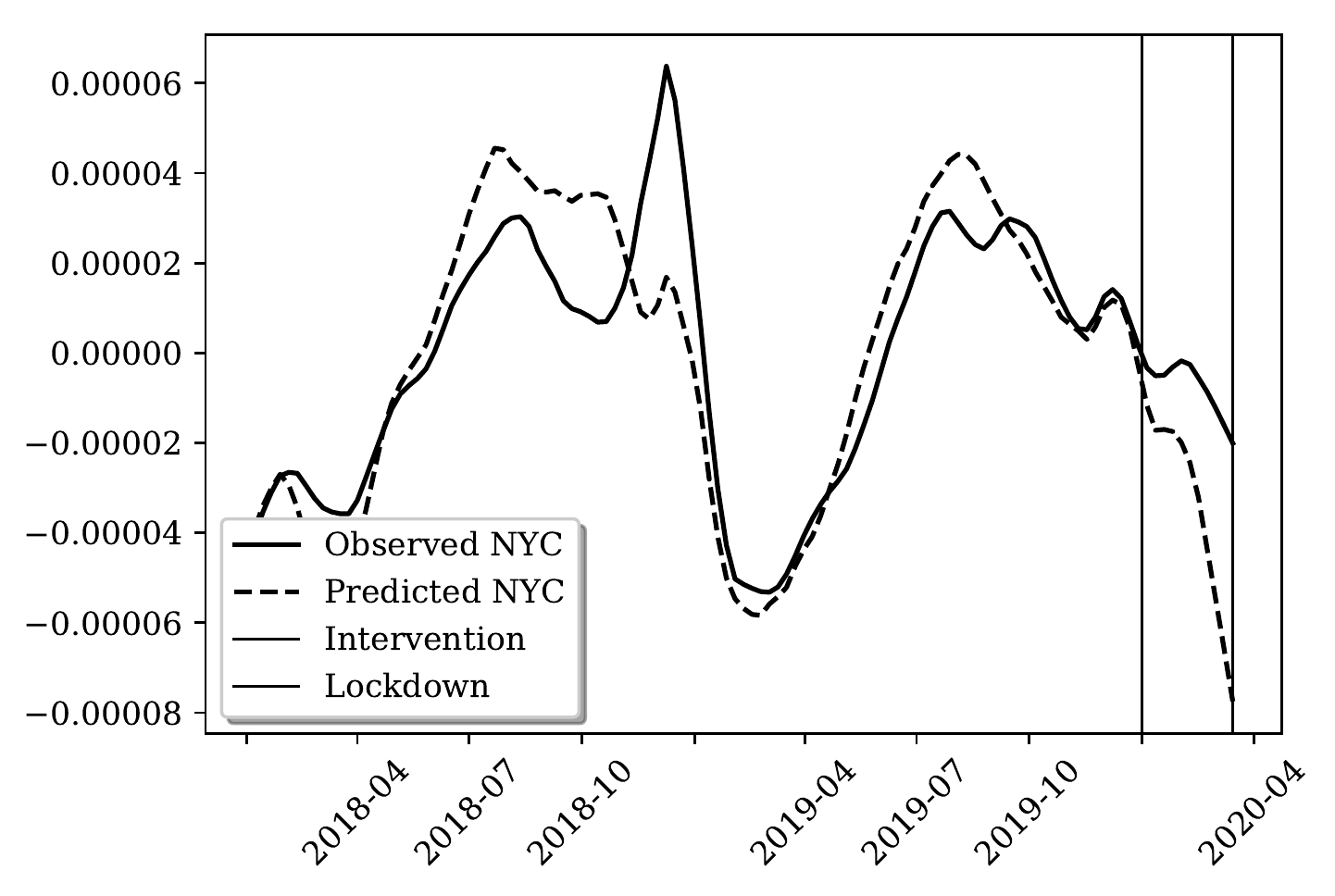}\caption{Property Theft}\end{subfigure}
\begin{subfigure}[t]{0.5\linewidth}\includegraphics[width=\textwidth]{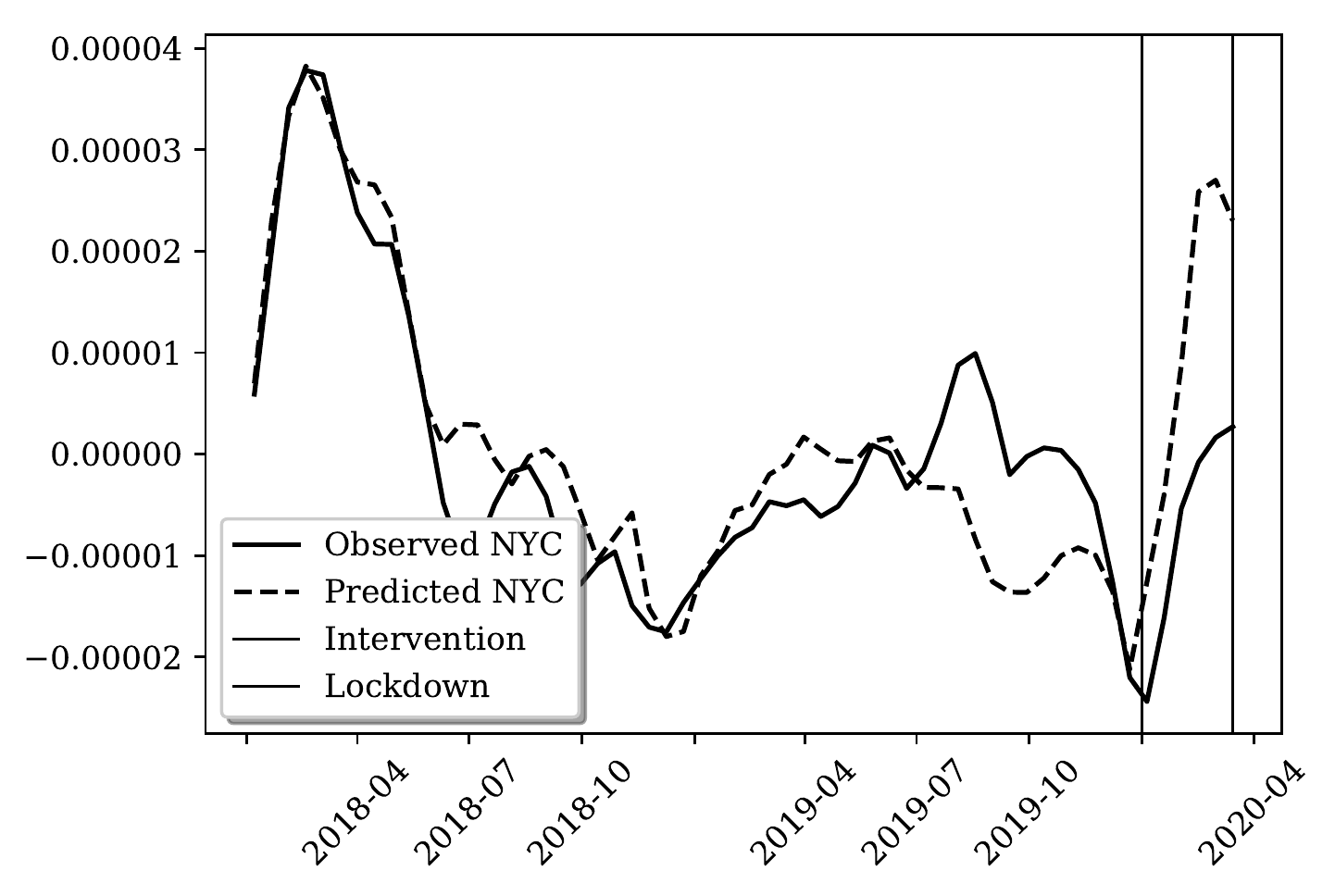}\caption{Drug}\end{subfigure}
\begin{subfigure}[t]{0.5\linewidth}
\includegraphics[width=\textwidth]{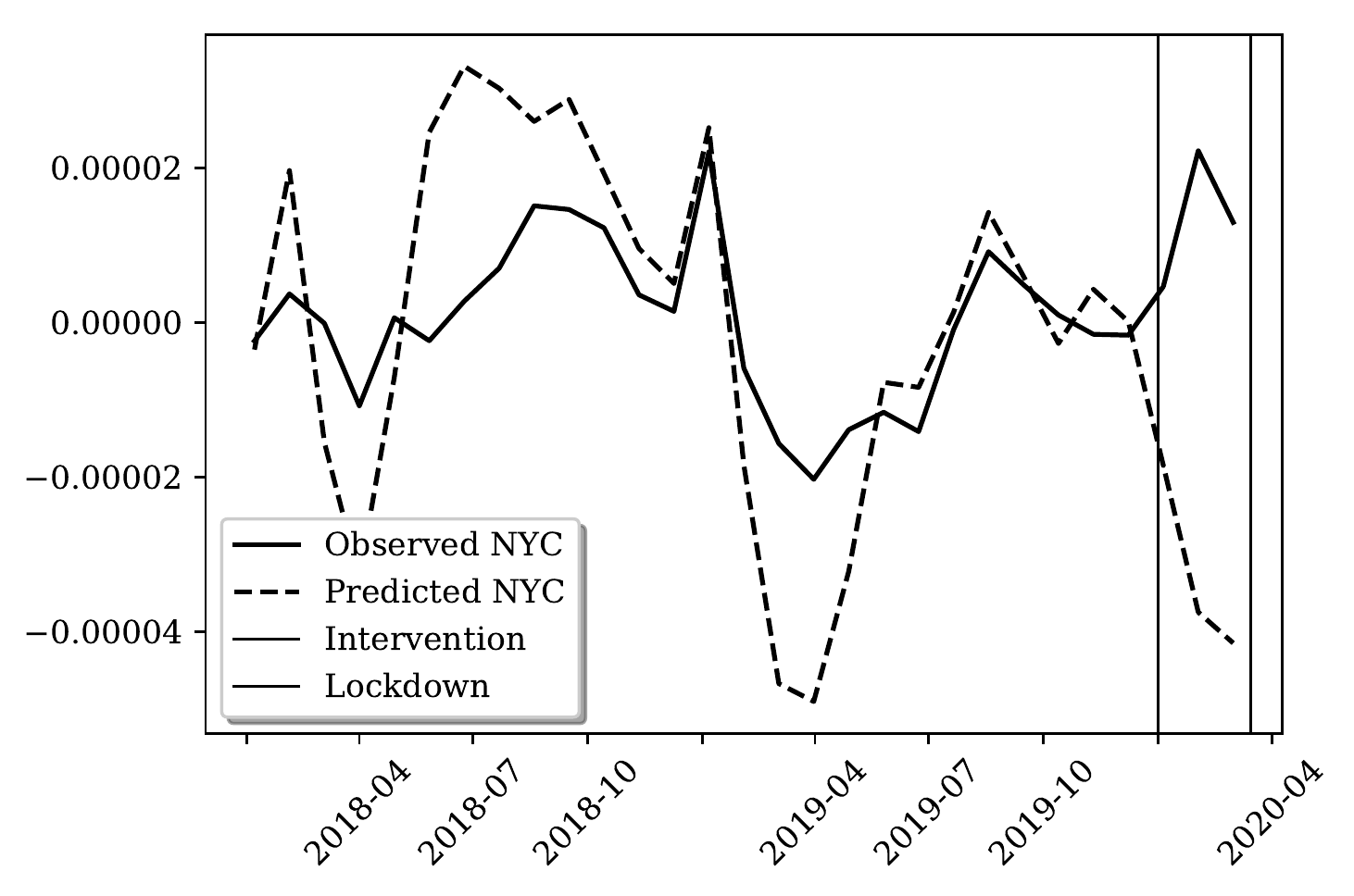}\caption{Property Burglary}\end{subfigure}

\begin{subfigure}[t]{0.5\linewidth}\includegraphics[width=\textwidth]{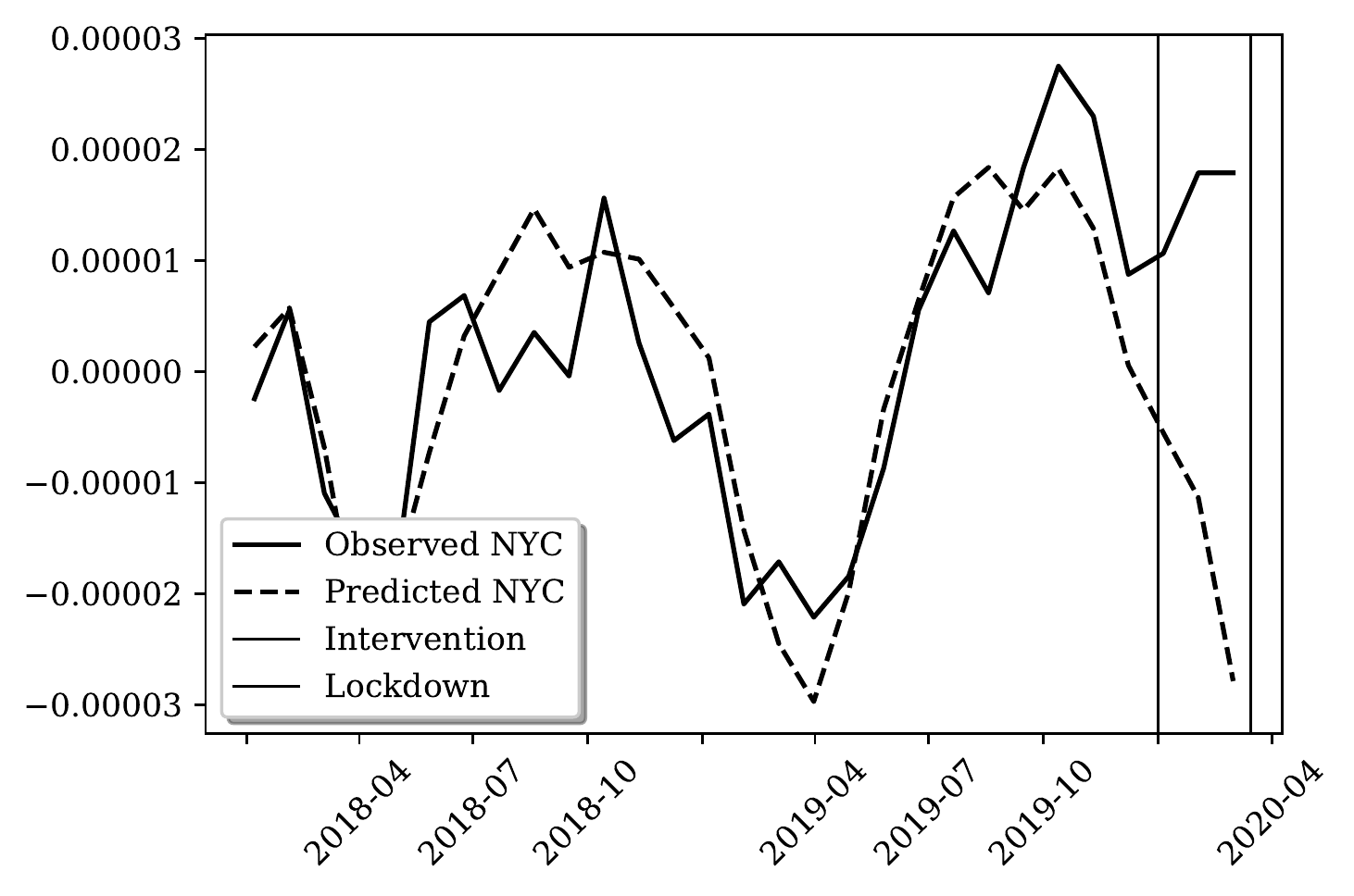}\caption{Violent Robbery}\end{subfigure} 
\caption{Synthetic control plots, Ridge (Loess-smoothed for display)}
    \label{fig:sc-visual-ridge}
\end{figure}

Our preferred specification is based on synthetic control via linear regression with a ridge penalty. 
 We include a visual comparison of the NYC time series, as well as the imputed synthetic control fit; visual overlay of the placebo fits and the NYC-treated fit, a summary table of ATE estimate and $p$-values, as well as summary statistics on pre-treatment fit, testing at level $\alpha=0.1$ adjusted for multiple-tests via Holm-\v Sid\'ak correction, as well as the output of synthetic control weights. Then we include tables of placebo checks: the in-time placebo checks and early roll-in of intervention placebo checks. We walk through this series of analyses for our preferred specification, and summarize the main inferential conclusions across specifications.

In \Cref{fig:sc-visual-ridge}, we show results for the synthetic control analysis for \textit{Assault} and \textit{Property Theft} outcomes, computed on a weekly aggregation, \textit{Burglary} and \textit{Robbery}, after aggregating event counts to a five-weekly coarsening. The other crime types are much lower frequency (ranging from 20\% to 5\% of the average frequencies as \textit{Assault}). Therefore, aggregating at the temporal level is required to reduce noise of the synthetic control estimation. For purposes of display, we graphically display loess-smoothed time series (using 7\% of the data to estimate each value). \edit{In the solid black line, we display the time series of observed cases. The vertical dark lines indicate the time of intervention  and COVID lockdown (end of study period), respectively. The dashed black line indicates the predicted NYC time series. The average treatment effect estimate differences the average of the post-and pre-treatment difference between these time series. The three crimes, assault, theft, and drug crimes, have good pre-treatment fit: the predicted time series closely matches the observed time series. The burglary and robbery time series are higher-variance and pre-treatment fit is not as close, but matches overall trends. For assault, theft, and drug, the post-treatment difference visually appears fairly close. For burglary and robbery, the post-treatment difference visually appears to increase, but due to the noise in pre-treatment fit, it's not clear whether this can be statistically distinguished from noise. Next we describe the treatment effects quantitatively. }

\begin{table}[t!]
\caption{Synthetic control (Ridge) tabular results}
\begin{tabular}{l|lllllll}
\hline \hline 
& ATE (/1000) & p (ATE) & p (RMSE) & Pre $R^2$ & Pre RMSE & Plac. RMSE & Num. plac.\\ \hline
 Assault&$0.0083$ & $0.35$ & $0.53$ & $0.76$ & $8.19 \times 10^{-6}$ & $2.38 \times 10^{-5}$ & $20$\\
 Theft&$0.0274$ & $0.18$ & $0.1$ & $0.64$ & $2.02 \times 10^{-5}$ & $4.54 \times 10^{-5}$ & $22$\\
 Burglary&$0.0458$ & $0.15$ & $0.37$ & $-1.32$ & $1.53 \times 10^{-5}$ & $2.45 \times 10^{-5}$ & $20$\\
 Robbery&$0.0304$ & $0.05$ & $0.0$ & $0.7$ & $7.4 \times 10^{-6}$ & $1.62 \times 10^{-5}$ & $20$\\
Drug&$-0.014$ & $0.71$ & $0.23$ & $0.7$ & $8.66 \times 10^{-6}$ & $2.34 \times 10^{-5}$ & $14$\\
\end{tabular}
\label{tab:sc-results-ridge} 
\end{table}

\begin{table}[t!]
\centering\caption{Lower and upper confidence bounds on effect sizes (rejected under the null) }
\begin{tabular}{l|ll}
& Lower bound (5th percentile) & Upper bound (95th percentile) \\ \hline 
 Assault&$-4.95 \times 10^{-5}$ & $2.01 \times 10^{-5}$\\
 Theft&$-5.95 \times 10^{-5}$ & $5.83 \times 10^{-5}$\\
 Burglary&$-1.54 \times 10^{-4}$ & $8.34 \times 10^{-5}$\\
 Robbery&$-5.47 \times 10^{-5}$ & $3.08 \times 10^{-5}$\\
Drug&$-7.95 \times 10^{-5}$ & $4.26 \times 10^{-5}$\\
\end{tabular}
\label{tab:sc-ridge-effectsizes} 
\end{table}

\begin{table}[t!]
    \centering
        \caption{Multiple testing adjustment, Ridge (for ATE, RMSE test statistics separately)}
    \label{tab:sc-multiple-testing-p-ridge}
    \begin{tabular}{l|llllll}
    \hline \hline 
& Robbery &  Theft &  Burglary &  Assault & Drug\\ \hline
Adjusted p (ATE)&$0.23$ & $0.48$ & $0.48$ & $0.58$ & $0.71$\\ 
Conclusion&Fail to Reject & Fail to Reject & Fail to Reject & Fail to Reject & Fail to Reject\\ 
Adjusted p (RMSE)&$0.0$ & $0.33$ & $0.54$ & $0.6$ & $0.6$\\ 
Conclusion&Reject & Fail to Reject & Fail to Reject & Fail to Reject & Fail to Reject    \end{tabular}

\end{table}
In \Cref{tab:sc-results-ridge}, we summarize the main numerical results of the synthetic control analysis, for each crime type. We report the averaged ATE estimate (on the event scale, normalized to be a rate per 1000 individuals). That is, the synthetic control comparison suggests that the increase in average weekly violent assaults is $0.0083$ events per 1000 individuals, relative to the synthetic control. The ``p (ATE)" column reports the $p$-value from the unit-level placebo distribution of the ATE estimate, while the ``p (RMSE)" column reports the $p$-value from the RMSE test statistic. We also report measures of pre-intervention predictive fit of the synthetic control in predicting observed outcomes: ``Pre $R^2$" is the $R^2$ value of the pre-treatment fit for the main synthetic control analysis where NYC is the treated unit. We also compare the pre-intervention RMSE (``Pre RMSE") and average placebo pre-intervention RMSE (``Plac. RMSE"). The last comparison, pre-intervention RMSE vs. average placebo pre-intervention RMSE, provides a sense of the predictive fits for the placebo synthetic controls. Finally we report the number of placebo units (``Num. plac.") that remain after the RMSE prescreening step. 

\begin{table}[ht!]
    \centering
        \caption{Synthetic control tabular results, Ridge, weekly aggregation}
    \begin{tabular}{l|lllll}
    \hline
    \hline
City& Assault &  Theft &  Burglary &  Robbery & Drug\\ \hline
Austin&$0.04$ & $0.05$ & $0.07$ & $0.07$ & $0.27$\\
Baltimore&$0.01$ & $0.05$ & $0.04$ & $-0.0$ & $--$\\
Boston&$0.07$ & $0.05$ & $0.07$ & $0.06$ & $-0.03$\\
Buffalo&$0.03$ & $0.02$ & $-0.02$ & $0.03$ & $--$\\
Chicago&$0.01$ & $0.05$ & $0.05$ & $0.04$ & $0.06$\\
Cincinnati&$0.01$ & $0.06$ & $0.0$ & $0.0$ & $--$\\
Dallas&$0.07$ & $0.07$ & $0.06$ & $0.05$ & $0.14$\\
Denver&$0.07$ & $0.02$ & $0.06$ & $0.05$ & $-0.11$\\
Detroit&$-0.01$ & $0.04$ & $0.03$ & $0.05$ & $-0.07$\\
Kansas City&$0.01$ & $0.01$ & $-0.01$ & $-0.0$ & $-0.01$\\
Little Rock&$0.03$ & $0.0$ & $0.01$ & $0.05$ & $--$\\
Los Angeles&$0.06$ & $0.08$ & $0.08$ & $0.06$ & $0.3$\\
Louisville&$0.03$ & $0.02$ & $0.05$ & $0.06$ & $0.01$\\
Milwaukee&$0.04$ & $0.04$ & $0.01$ & $0.03$ & $--$\\
Nashville&$0.04$ & $0.06$ & $0.01$ & $0.01$ & $0.11$\\
Philadelphia&$0.05$ & $0.05$ & $0.06$ & $0.06$ & $0.04$\\
Phoenix&$0.09$ & $0.07$ & $0.07$ & $0.06$ & $0.15$\\
Portland&$0.06$ & $0.05$ & $0.06$ & $0.06$ & $0.18$\\
Raleigh&$0.07$ & $0.06$ & $0.06$ & $0.06$ & $-0.05$\\
Seattle&$0.06$ & $0.05$ & $0.06$ & $0.06$ & $0.02$\\
Virginia Beach&$0.06$ & $0.05$ & $0.08$ & $0.07$ & $0.02$\\
Washington DC&$0.09$ & $0.03$ & $0.07$ & $0.05$ & $--$\\
intercept&$-0.0$ & $0.0$ & $0.0$ & $0.0$ & $-0.0$\\
\end{tabular}
\label{tab:sc-ridge-weights} 
\end{table}

To put the effect sizes in context, the ATE point-estimate in number of assault events is 69.9 events relative to approximately 1760 weekly assault events; or an estimated 3.9\% relative percentage increase. For Theft, the ATE point-estimate is 230.7 events relative to approximately 2667 weekly events, or 8.7\% relative increase. {To convert our per-capita estimates to absolute events, we obtain this estimate by computing $0.0083/1000 * \text{2020 NYC population}$.} Further, to characterize the extremal boundaries of effect sizes that would result in failing to reject the null hypothesis, we report lower and upper bounds from the placebo test distribution in \Cref{tab:sc-ridge-effectsizes}. 
We also report multiple testing adjustments via the Holm-\v Sid\'ak correction in \Cref{tab:sc-multiple-testing-p-ridge} which adjust for the probability of rejecting the null by chance over multiple comparisons.

For property theft, robbery, and drug, there is some sensitivity to conclusions depending on the test statistic (although with a multiple testing-adjustment via Holm-\v Sid\'ak correction reported in \Cref{tab:sc-multiple-testing-p-ridge}, at a $\alpha=0.1$ level we only reject the null for the violent robbery crime type). For burglary, we have the most consensus between the ATE and RMSE test statistics of non-significance of the increased effect.

For interpretation,
in \Cref{tab:sc-ridge-weights}, we display the weights chosen by the synthetic control. Note that we allow for a degree of extrapolation by allowing for negative weights.

\begin{figure}[ht!]
    \centering
    \begin{subfigure}[t]{0.5\linewidth}
    \includegraphics[width=\textwidth]{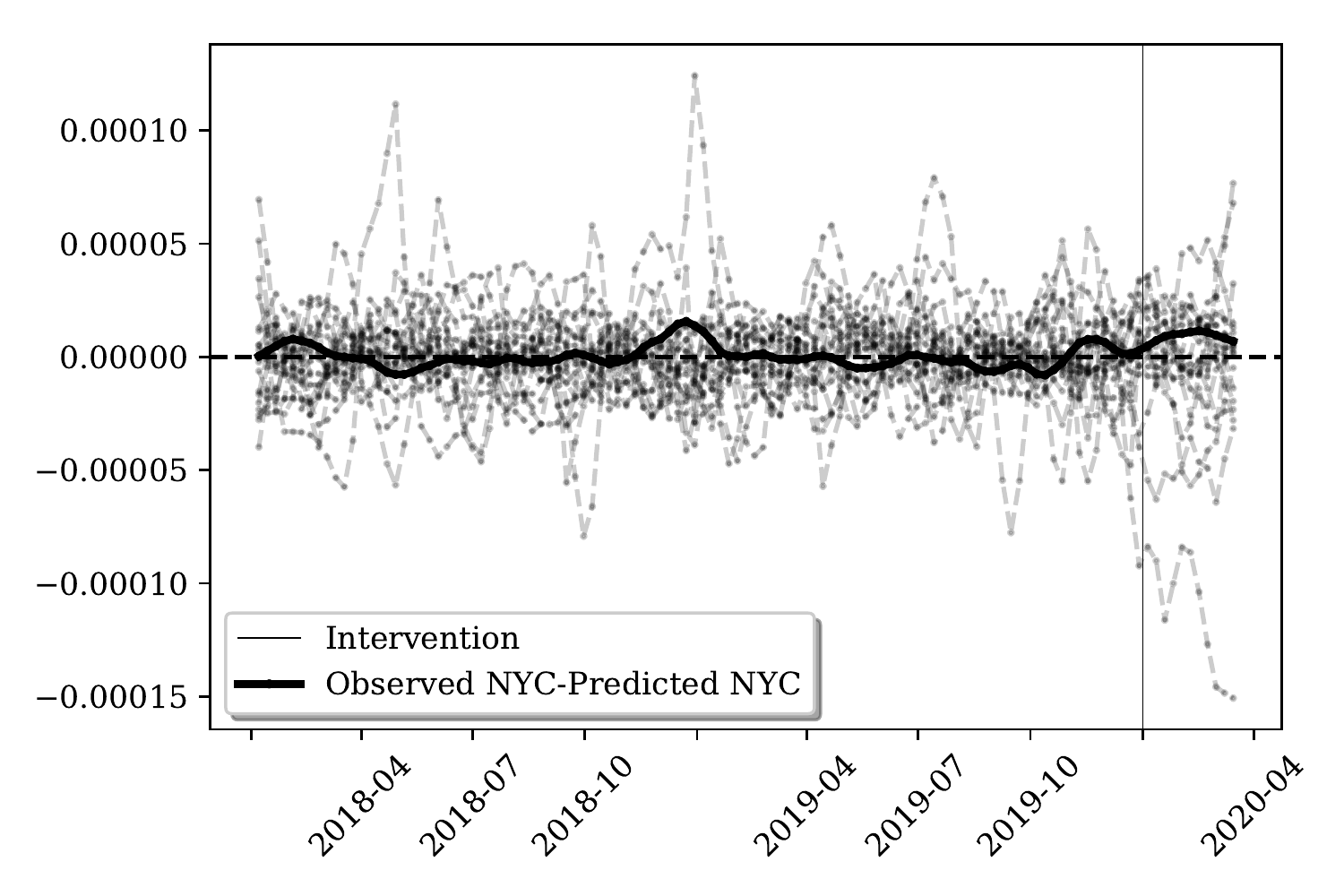}\caption{Violent Assault}\end{subfigure}\begin{subfigure}[t]{0.5\linewidth}
\includegraphics[width=\textwidth]{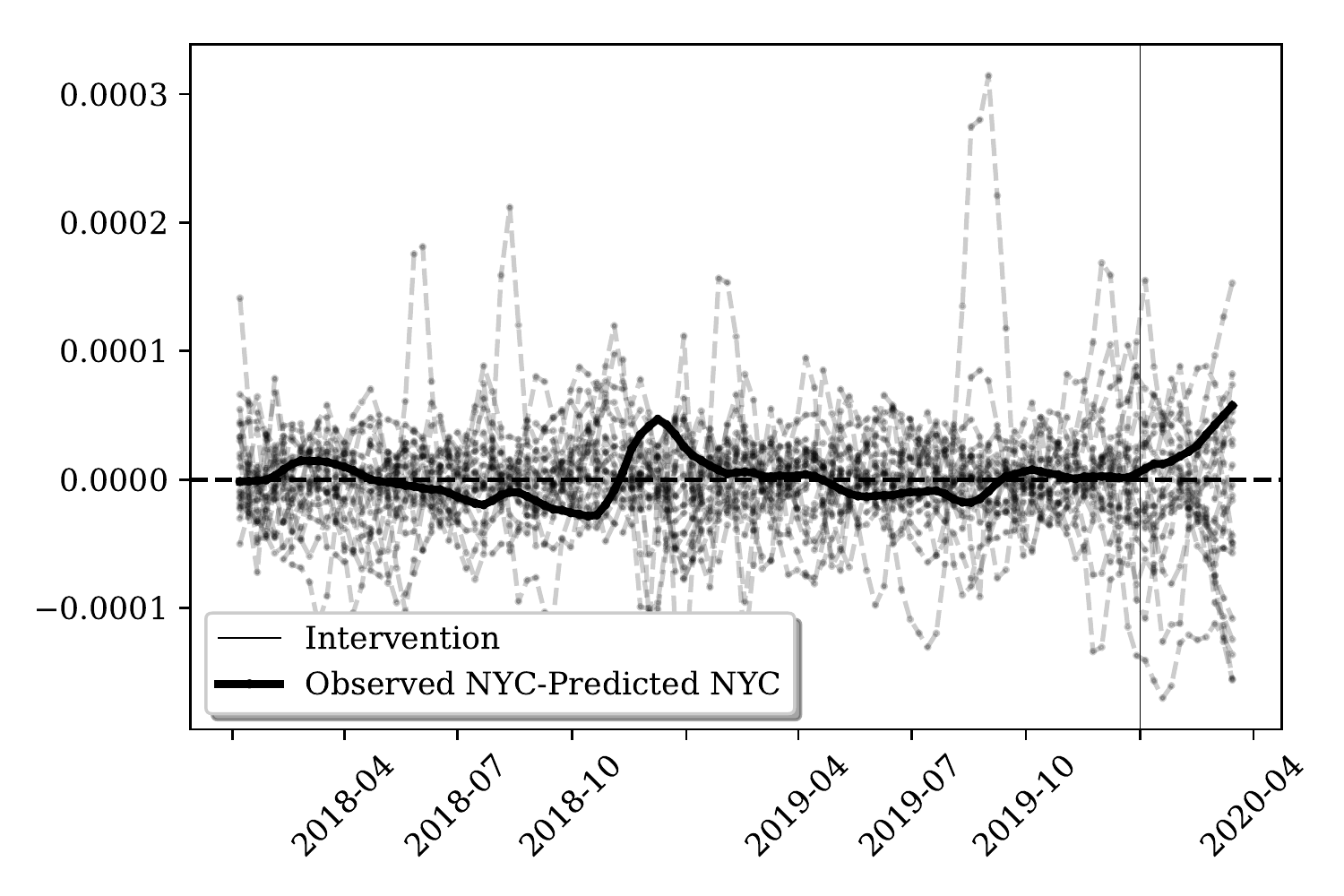}\caption{Property Theft}
\end{subfigure}
\begin{subfigure}[t]{0.5\linewidth}
	\includegraphics[width=\textwidth]{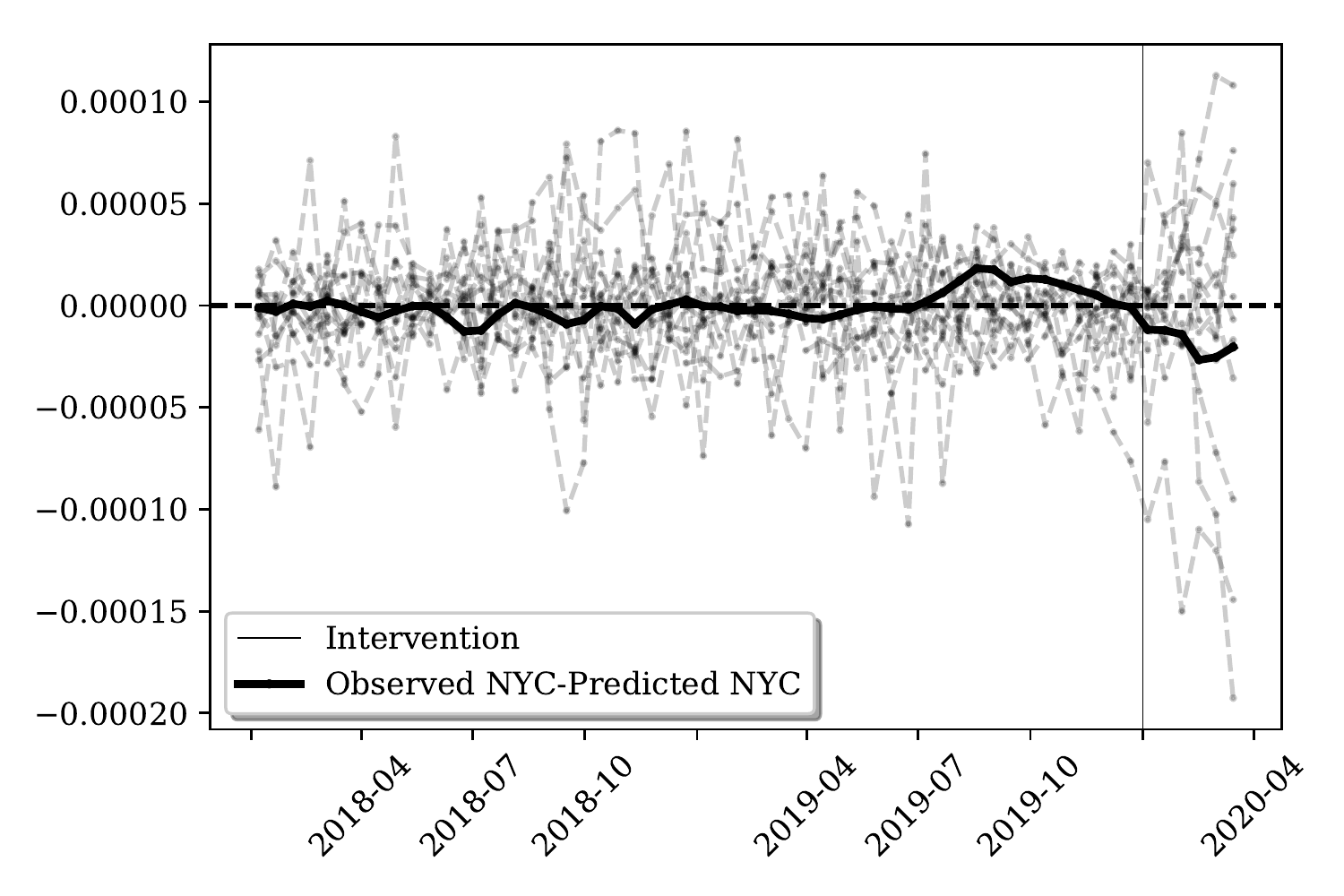}\caption{Drug}
\end{subfigure}
\begin{subfigure}[t]{0.5\linewidth}
      \includegraphics[width=\textwidth]{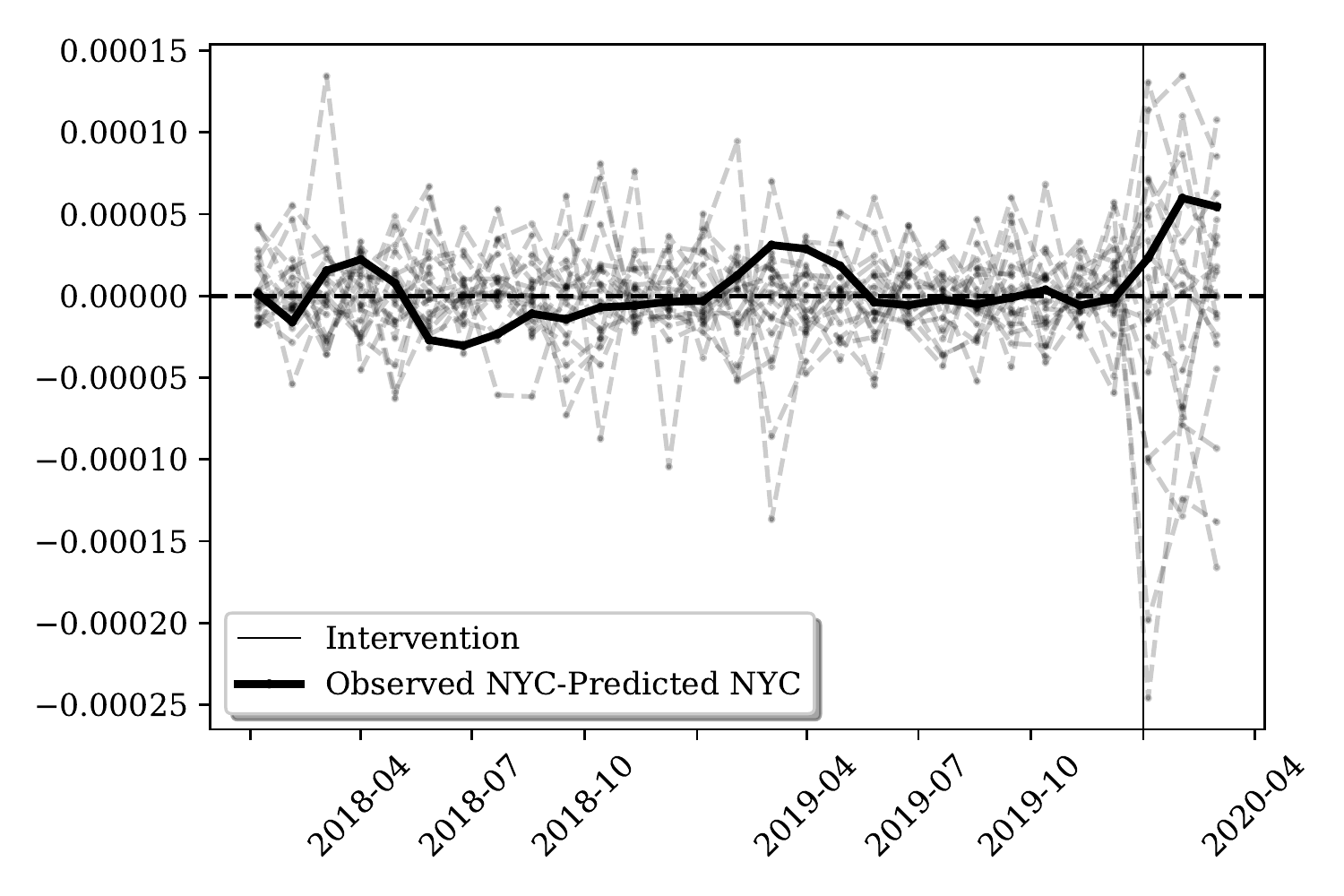}\caption{Property Burglary}\end{subfigure}\begin{subfigure}[t]{0.5\linewidth}
      \includegraphics[width=\textwidth]{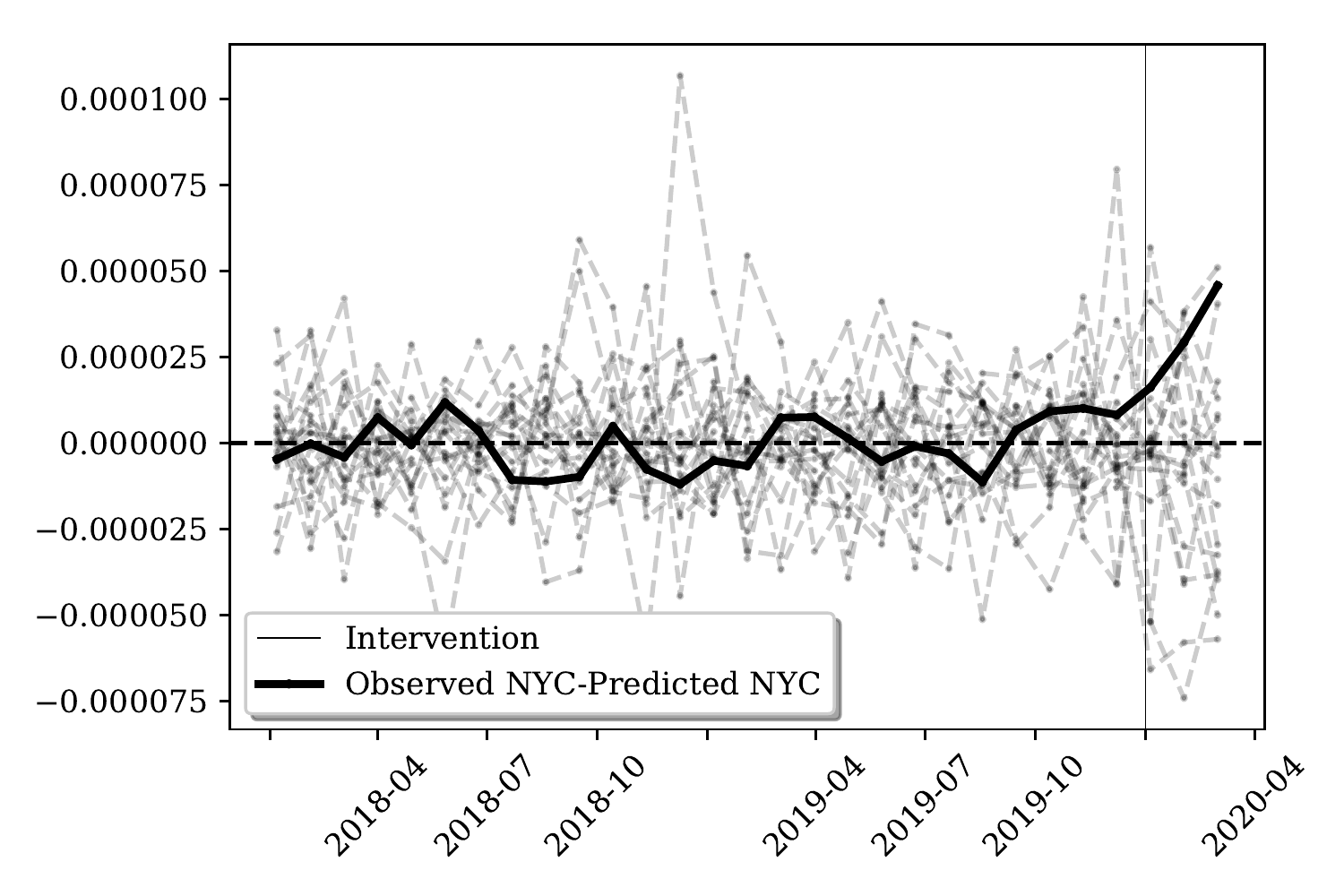}\caption{Violent Robbery}
      \end{subfigure}
    
    \caption{Unit-level placebo checks (Ridge; Loess-smoothed for display)}
    \label{fig:sc-visual-ridge-placebos-unit}
    \vspace{20pt}
\end{figure}
To summarize the findings from \Cref{tab:sc-results-ridge,tab:sc-multiple-testing-p-ridge}: qualitatively, we would conclude that significance testing at $\alpha=0.1$ would result in failure to reject the null hypothesis of no increase in crime, except for violent robbery. \edit{That is, for most crimes (except violent robbery), we do not find statistically significant evidence of an increase in aggregate crime.}

In \Cref{fig:sc-visual-ridge-placebos-unit}, we graphically display the placebo distribution and the NYC residuals in the context of the generated leave-one-out placebo distribution. For each city acting as the treated unit, we plot the residual series: actual observations minus the predicted values. This provides a qualitative sense of the validity of the placebo distribution. While the placebo fits tend to be noisier than the original NYC fit, the average RMSE is of the same order of magnitude. Comparing Violent Assault and Property Theft suggests that post-intervention deviations for the unit-level placebo checks are overall of similar magnitudes to the NYC-treated analysis.

We conduct a placebo check by assigning each other city as the treated unit, then recomputing a synthetic control from the other cities. We follow the suggestion of \citet{abadie2011synth} to compute the distribution of placebo effects, restricting attention to those cities where good pre-treatment fit is possible, relative to the original synthetic control. We consider two-sided tests of the weak null. 

One test statistic that we assess via the placebo test is the \textit{average treatment effect estimate} (ATE) defined in \cref{eqn:ate}.  
We then consider the $p$-value of the ATE measured for NYC with respect to the distribution of ATEs from the placebo checks. Namely, we set the $p$-value to $1-F(\hat\tau)$, where $F$ is the cumulative distribution function of the placebo ATE estimates and $\hat\tau$ is the ATE estimate measured for NYC. 

Another test statistic is the RMSE (root mean-squared error) statistic used in \citet{abadie2011synth}. 
The statistic $r$ is given by the ratio of post-treatment RMSE and pre-treatment RMSE: 
\begin{equation}
    r = \frac{ \frac{1}{T-T_0}\sqrt{\sum_{t=T_0+1}^{T} (Y_{1t}-\hat Y_{1t})^2 }   }{
    \frac{1}{T_0}\sqrt{\sum_{t=0}^{T_0} (Y_{1t}-\hat Y_{1t})^2 }
    }\label{eqn:rmse} 
\end{equation}
We can similarly compute placebo RMSE statistics by permuting the identity of the treated unit and the $p$-value as the cumulative distribution function of the primary RMSE statistic among these placebo values. 
Relative to the more interpretable ATE test statistic, the RMSE test statistic normalizes by the pre-treatment fit under a certain placebo. If a placebo unit, when considered as the treated unit, achieves a poor pre-treatment fit, then any extreme differences in post-treatment contrasts, which may rightly be attributed to the poor pre-treatment fit,
are made less extremal in the placebo distribution by the normalization.

In \Cref{sec:apx-robustnesschecks}, we include additional robustness and placebos in time. 

Overall the results of the placebo check suggest that under alternative specifications where we substantively know there is no treatment effect, we recover similar order of magnitude ATE estimates as under our main specification. Robbery does have some extremal $p$-values under the placebo check: it may suggest that the data series is quite noisy such that known null effects may nonetheless be extremal. Otherwise however, no statistical significance would be detected under the placebo specifications. The same conclusion holds under the early-roll-in of intervention start placebo check. Notably, despite the documented increase in ROR rates as we moved the intervention date earlier in time, we do not detect significant or extremal effects. 

\edit{In conclusion, using synthetic control methodology that is more approprirate in this setting shows overall null effects of bail reform on aggregate crime. Synthetic control can account for unobserved confounders and anticipations of policy changes. }

\section{Limitations}\label{sec:limitations}  
Our analysis of the NYC bail reform, conducted via aggregating police department-level incident reporting data, with a short post-intervention timespan before the onset of COVID, is subject to important limitations. Below, we describe these with relevant descriptive statistics and conclude, when possible, how these limitations may impact our estimates. We describe limitations due to studying general vs. specific deterrence, the short study period, a one-time retroactive release of defendants that occurred, and possible anticipation effects. 

\paragraph{The proportion of cases directly impacted by bail reform is small.}

The primary way bail reform could impact crime is through the release of defendants who would’ve otherwise been detained pretrial (ie, bail not paid/remanded). Between January 1, 2020 and March 15, 2020, there were a total of 18,485 cases with defendants who were (1) continued at arraignment, (2) ineligible for bail (flagging eligibility based on charge), and (3) not bailed/remanded at arraignment.

Though these cases were ineligible for bail, we need to further estimate the percentage that would have been bailed/remanded had the new bail law not been in effect. During the same period in 2019, 16.3\% of defendants were bailed/remanded in cases that met the conditions above (excluding condition 3, not bailed/remanded). 55\% of these defendants were eventually released back into the community during the pretrial period. This matters if we’re thinking about the marginal difference in terms of time at risk in the community, since after a release there would be little difference between these cases and cases mandatorily released post-reform. Applying this percentage to 2020 yields: 3,013 cases that would otherwise have been bailed/remanded. This small proportion of cases impacted could lead to a downward bias of our estimate since NYC was already releasing a large percentage of defendants, so the number of defendants on the margin who were affected by the reform was low.

\paragraph{Short study period.}

The post-intervention period in our study runs from January 1, 2020 to March 15, 2020 – only a 75-day period.

Among time to disposition of new incidents, and the amount of time on release in the community, the window with potential for rearrest in our post-intervention period is small. 

In NYC, speedy trial rules require that misdemeanor cases be resolved within 60-90 days and felony cases within 180 days. However, we know the typical time to disposition is much longer (eg, in 2019, the average time to disposition in indicted felony cases was ~10 months \citep{weill2021felony}). On the other hand, during the 75-day study period, the number of days a case could have been on release in the community ranged from 1 to 75, with an average of about 30 days (assuming an equal distribution of cases for each day during the 75-day period).

This means that for the roughly 3,000 defendants that would have otherwise been bailed/remanded, the study only captures a small portion of their overall time at risk (3,000 * 30 days = 90,000 person-days of risk). In 2020, the pretrial rearrest rate for cases tracked through March 2021 was about 20\% according to Office of Court Administration, \citep{pretrial_release_data_OCA}. Given the short time at risk in the study (30 days on average), we need to assume a much lower rate of rearrest relative to statistics derived from longer time periods. For the sake of argument, suppose the pretrial rearrest rate from our shorter time period was 10\%, so 300 of the 3,000 defendants impacted by the reform were rearrested by March 15, 2020. On the other hand, the clearance rate during the study period was about 1 in 3 \citep{nypd_clearance}. A worst-case analysis, assuming these individuals were responsible for 3x the number of rearrests, suggests 900 rearrests in total. Overall, the impact of a shorter time frame for the study suggests that our estimates could be biased downwards from the estimate with a longer time-period, because of the short time at risk for re-offense.

\paragraph{Seasonality.}

The study occurs entirely during the winter months when crime is at its lowest ebb. While the models account for temperature and month-week fixed effects, we may not have enough historical data to identify the true seasonal patterns. This could bias estimates downwards because the post-intervention period is entirely within a seasonal lull in crime.

\paragraph{One-time retroactive release of defendants.}

It was estimated that about 900 defendants were retroactively released under the new law \citep{onetimerelease}. Our estimates could therefore overstate the effect of the policy, since during the study period there were many more impacted defendants (released) than there would normally have been. Our early intervention placebo checks verify that the null results we observe are not due to a cancellation in positive treatment effect. If there \textit{were} a positive treatment effect, due to this large retroactive release in November, a concern is that using our later intervention date wrongly suggests a null. But in our early intervention roll out placebo checks, we verify that these would have remained null findings, addressing this counterargument.

\paragraph{Anticipation effects.}

Many of the retroactive releases (described above) occurred ahead of the policy start date. Also, there is some evidence that judges had already started to change their behavior in anticipation of the new law in November-December, 2019 (less likely to bail/remand defendants who would soon be released anyway). Although this could lead to a downward bias in estimates, this is addressed in our early intervention roll-in sensitivity analysis, which also assesses null effects. The earlier sensitivity analysis suggests that this is not a serious limitation.

\section{Conclusion}\label{sec:conclusion}

This paper studies the impact of bail reform on aggregate crime, using the case study of New York's 2020 reforms. 
Subject to the limitations discussed above, we find evidence that bail reform did not increase aggregate crime rates in NYC.

We consider a subset of crime outcomes, Assault, Theft, Drug, Burglary, and Robbery, which are the most frequent time series. 
Our main analysis uses the synthetic control method, which allows for adjusting for cross-unit variation by comparing NYC's trend to a synthetic control, constructed by weighting the time series of other cities. In order to provide some sense of uncertainty quantification, we conduct a series of placebo checks as conventional in the synthetic control literature, and report $p$-values from the placebo distribution of ATEs that arise from fixing another unit as the treated outcome, to assess the variation in null effects arising from synthetic control.
The results from the synthetic control overall fail to reject a null hypothesis of no effect for most outcomes (before and after adjusting for multiple comparisons); with the exception of rejecting the null (finding a statistically significant increase) for the robbery outcome. Our analysis focused on assessing the impacts of bail reform on aggregate crime rate, hence informing general deterrence considerations of the broader policy context. Overall, we find evidence that bail reform did not increase aggregate crime rates in New York City.

\bibliography{synth}
\bibliographystyle{apalike}

\appendix
\counterwithin{figure}{section}
\counterwithin{table}{section}

\clearpage
{
\begin{center}
\Large{\textbf{Supplementary material}}
\end{center}
}
\paragraph{Table of Contents} 
\begin{itemize}
\item \Cref{sec:apx-addldiscussion} includes additional discussion omitted from the main text. 
\item \Cref{sec:apx-datapreprocessing} discusses data preprocessing steps used in the method. 
    \item \Cref{sec:apx-its} includes ITS analysis and results. 
    \item \Cref{sec:apx-robustnesschecks} includes additional robustness checks.
    \item \Cref{sec:apx-datasetdescription} contains a more detailed description of the dataset.  
\end{itemize}
\section{Additional Discussion}\label{sec:apx-addldiscussion}

\subsection{Crime trends}
We first discuss the backdrop of underlying national crime trends, against which any observed increase in crime would have to be compared in order to draw qualitative conclusions attributing increases to the bail reform intervention itself. Such national crime trends reflect observed and unobserved confounders affecting adoption of bail reform and crime. We review the criminological literature which discusses these shared confounders: although there is qualitative consensus for the relevance of overall national trends, such trends may arise from a number of mechanisms and hence are hard to measure directly. The backdrop of national crime trends is important for assessing changes in crime rate and understanding challenges in disentangling national trends from effects of the bail reform in New York. In the context of our analysis, it provides conceptual grounding for why statistical methodology that accounts for shared unobserved confounding can be relevant in our setting. Although crime is largely a local phenomenon \citep{oct21,cw13,s08}%
, there are overall national trends that emerge from a combination of various micro and macro level factors \citep{b10}. For example, the great crime decline that began in the 1990s and continued well into the current century was likely effected by an interplay of an expanding economy, increased incarceration and more effective policing, demographic shifts \citep{b10,z07,b08,bw06,dl01,d17}%
, urban development \citep{s99,bcmtjt11}%
, changes in youth culture \citep{c98,o02}%
, shifts in illegal drug use and distribution \citep{d17}%
, immigration \citep{re07,s08}%
, improved security technology \citep{ftt14,op15}%
, and the phasing-out of lead paint and leaded gasoline \citep{cn10,n07,r02,sl04,n04,g09,sw18}%
. However, there is no consensus on the most important causes or exact mechanisms behind the crime drop \citep{b10}%
. Moreover, certain explanations proposed by some scholars have been disputed by others, such as the significance of the role of incarceration and policing \citep{z07,rlb15,o02}
or the impact of legalized abortion \citep{dl01,br08,b08,o02}.

Similarly, there is a lively debate over the causes of the unexpected increase in violent crime that started in 2015 \citep{r16,s18,j18,d17}%
 or the sharp uptick in murder in 2020 \citep{c21}%
. In contrast to the 1990s crime drop, these recent trends are unique to the United States \citep{wstk16,ftt14,o02,j21,l21}%
 and limited to violent offenses, particularly homicide, while property crimes continued to decline \citep{donohue2019right,rw19,rl20}.%

There is a general notion that the COVID-19 lockdowns and widespread social unrest triggered the rise in violence \citep{j21,cassell2020bail,l20}%
, but it is possible that the unusual events of 2020 merely exacerbated an upward trend that has been in the making for a while \citep{d17}. That is, the recent reversal of the downward trajectory in violent crime may be due to a combination of several longstanding trends, including a drop in incarceration rates, declining police rates, emerging illegal drug markets, and an increase in out-of-wedlock births. Even though these patterns have emerged gradually over several years, they could have ultimately had a discontinuous impact on violent crime \citep{d17}%
. The seemingly paradoxical fact that property crime rates continued to fall may be the consequence of the strong economy, technological progress (for example, cashless payment methods and the lower cost of goods reduced the incentives to commit robberies or burglaries), or underreporting \citep{d17}%
. 

To be sure, the pandemic was accompanied by several developments that could be plausibly linked to crime, such as increased economic hardship \citep{pcmww20}%
, decarceration \citep{p21}%
, an uptick in gun purchases \citep{smpstablw21}%
 and gun carrying \citep{aa21}%
, and halted non-policing efforts to prevent violence \citep{p20,bb20}%
. However, except for one study, which found a positive association between the recent increase in firearm sales and gun violence \citep{smpstablw21}%
, thus far there is no rigorous empirical evidence that these factors caused an additional spike in homicide rates. 

Likewise, the impacts of the Floyd protests on crime rates are still unclear. Despite the perception that defunded police forces led to the crime increase \citep{efj21,y21}%
, police budget cuts were in fact only modest and not universal \citep{ahc21}%
. Moreover, violence also went up in jurisdictions that increased their police spending \citep{j20,ab21,a21}%
. Public trust in law enforcement eroded in the wake of police misconduct and social unrest
\citep{j20,rn21}%
, which could have suppressed proactive policing and in turn increased crime \citep{mbdsm13,dpk16,gr08,rw19,c21}%
. However, it has yet to be shown whether this was the case in 2020. 

In summary, despite the controversies and uncertainties surrounding the exact causes behind broad shifts in crime incidence, there are undoubtedly forces at play that affect jurisdictions across the country in similar ways.

On the other hand, criminological background on mechanisms of crime, such as crime specialization, can inform our assessment of the impact of New York's bail reform, which changed the incidence of bail for some crimes more than others. Differences in charge eligibility across crime types, in combination with implications of the research on crime specialization, could predict heterogeneity in possible effects of bail reform on different types of crime. Research on crime specialization seeks to understand whether people tend to commit different types of crimes, for example during life course analysis. \citet{macdonald2014linking} note that recent research suggests ``that there are meaningful differences in the tendency for criminally active persons to repeat violent offenses compared to other offense types". Applying this mechanistic insight to this treatment effect evaluation context, we might have expected that if bail reform led to a situation where more defendants who were charged with a violent felony were now on pretrial release, then we might observe more violent offenses, due to observed patterns of crime specialization. However, as described above, bail reform did not affect eligibility for pretrial release for violent felony charges, although it led to major changes in eligibility for non-violent felonies and misdemeanors. 
\subsection{Bail reforms elsewhere}\label{sec-apx-addldiscussion-bailreforms}
We discuss more specifically the timing of bail reform policies among the control units. Cook County, which contains Chicago, passed General Order 18.8A (GO18.8A) on September 17, 2017, which required Cook County bond court judges to maintain a presumption against requiring the defendant to pay money to achieve release \citep{judge2019}. 
Harris County (which contains Houston) mandated pretrial release of a subset of misdemeanor offenses in 2019. San Francisco announced they would no longer seek cash bail for a subset of misdemeanor offenses; this went into effect late January 2020. Although Philadelphia instituted bail reform in February 2018, relative to these other cities, we instead treat their reform as a separate treatment from New York's reform. Philadelphia underwent prosecutor-led bail reform in February 2018, near the start of the study period. Unlike New York's bail reform which removed the ability of judicial discretion to set money bail, the prosecutor-led reform in Philadelphia still allowed discretion for bail magistrates to set bail. 
Unlike San Francisco and Houston, whose treatment dates occur in the middle of the study period, Philadelphia's bail reform occurred right at the beginning of our study period (going into place on February 21st, 2018), so a little less than $90\%$ of our study period is post-Philadelphia's bail reform. Philadelphia's bail reform was a prosecutor-led reform where the applied to a large fraction of cases in Philadelphia, there was discretion involved in whether bail magistrates set bail. \cite{ouss2020bail} find that
the No-Cash-Bail policy did lead to a ``sharp 22\% (11 percentage point) increase in the likelihood of being granted release on recognizance
(ROR, or release without monetary or supervisory conditions)", but not an increase in overall detention rates. In contrast, the original bail reform in New York, for most offenses, removed judicial discretion by completely removing the option of money bail for its noted offenses.  

Our setting poses a few challenges for direct application of models incorporating staggered adoption and synthetic controls. {Bail reform differs from the ideal case of staggered adoption that admits design-based analysis because arguments that appeal to design leverage, for example, quasi-random time-staggered rollouts due to implementation quirks, are less applicable to bail reform. And, there are multiple, different versions of treatment since the general terminology of ``bail reform" refers to different policies and policy settings. Besides the stated scope of the reforms, which is amenable to quantification, there are salient differences in policy regimes and implementation environment that are qualitatively important.}  
\clearpage

\section{Data preprocessing}\label{sec:apx-datapreprocessing}
We summarize data-processing degrees of freedom that arise with the synthetic control method here, including their data-driven or other justifications. 
\begin{itemize}
    \item \textbf{Tuning the ridge penalty by pre-treatment prediction quality (RMSE).}
Our preferred specification follows suggestions of \citet{abadie2015comparative,doudchenko2016balancing} by adding a ridge penalty. \cite{doudchenko2016balancing} show how in the absence of perfect pre-treatment fit, penalty terms on the weight norm can improve statistical precision while other constraints are dropped from the original specification (such as allowing negative weights). Overall, these modifications improve pre-treatment fit and statistical precision in the absence of perfect pre-treatment fit, at the expense of a small amount of extrapolation bias. The optimal penalty is determined separately for each outcome type based on optimal pre-treatment predictive fit. %

We still require the weights to sum to 1. We range the penalty, $\lambda$, on a logarithmically spaced grid of 100 values from $10^{-8}$ to $10^{-2}$. (We determined the grid size by expanding the boundaries until the optimal $\lambda$ penalty value was in the interior.) We use the first 80\% of pre-treatment data for learning the weights and choose $\lambda$ based on best (pre-treatment) performance on the held out set; the next 20\% of pre-treatment data. 
\item \textbf{Determining aggregation level for outcome series. }
We analyze Assault and Theft outcomes on a weekly aggregation, Drug on a bi-weekly (every two weeks) aggregation, and Robbery and Burglary on a tri-weekly (every three weeks) aggregation. 

We determined these weekly aggregations by, for the Drug, Robbery, and Burglary outcomes, stepping up to higher levels of aggregation until a satisfactory pre-treatment fit was achieved (and increasing aggregation level led to marginal improvements in pre-treatment fit). This is analogous to the use of ``scree" plots in choosing parameters in other statistical methods. Such a method for outcome-specific temporal aggregation was also used in \cite[Table 1]{robbins2017framework} because different crimes have different frequencies.  

This is because the individual time series are highly variable for other crimes (burglary, homicide, rape, and robbery), to the extent that the signal-to-noise ratio is quite low. When the pre-treatment fit is very poor, the synthetic control is uninformative --- even a $p$-value that does not reject the null hypothesis could otherwise be explained away by highly variable pre-treatment fits on the placebo synthetic controls, despite a qualitative difference in the predicted and observed time series. 

\item \textbf{Omitting cities based on pre-treatment fit more than 7.5 times that of the original synthetic control.} \citet{abadie2011synth}'s survey suggests to omit units from the placebo distribution with poor pre-treatment fit. Following this suggestion, we therefore, for every outcome, compute the placebo distribution from cities with pre-treatment fits no more than 7.5 times that of the original synthetic control. 

We argue that including units with extremely bad synthetic control fits can add extremal values to the placebo distribution that are not facially valid (because of an extremely poor synthetic control pre-treatment fit), and so could make it more likely to reject the null hypothesis simply on the grounds of including additional units where there is no valid synthetic control fit. Therefore, dropping these units reduces the variability of the placebo distribution, hence making Type-II errors less likely (failing to reject the null-hypothesis when it was actually false).

\item \textbf{Removing cities due to data reporting changes.}
Some, but not all cities, reported UCR codes which were used when available. Austin, Chicago, Cincinnati, Dallas, Los Angeles, Philadelphia reported UCR codes; but not all at the same granularity (with some reporting coarse codes at the hundreds level). Houston's schema changed twice during the study period; crime counts were discontinuous across schema changes under the aggregations and so we excluded Houston from the analysis. 

We removed Atlanta and Fort Worth because of data quality reporting issues: due to changes in reporting scheme, the observed time series has a large discontinuity. Fort Worth and Houston both moved to National Incident-Based Reporting System (NIBRS) reporting in 2018 which aligns with the anomalies for those cities. Kansas City also moved from encoding with Uniform Crime Reporting (UCR) codes to NIBRS descriptions in 2019; there also appears to be a data changepoint in the series in that time range.
\end{itemize}

\section{Interrupted-time-series analysis}\label{sec:apx-its}

\subsection{Interrupted-time-series analysis} 
As a starting point for the analysis, we conduct an \edit{Interrupted-time-series (ITS) analysis} \citep{bernal2017interrupted}. Due to autocorrelation of the time series and seasonal effects, we adjust for autocorrelation via segmented regression with autoregressive integrated moving average models (ARIMA) time series regression \citep{brockwell2016introduction}. We first adjust for seasonality via seasonal fixed effects and consider ARIMA$(p,d,q)$ models, where $p,d,q$ are positive integers. These denote: $p$ the order of the autoregressive (AR) part of the model, $d$ the degree of backwards differencing to achieve stationarity, and $q$ the order of the moving average (MA) part of the model. 

In an ARIMA model, $Y_t$ satisfies a difference equation of the form 
$$ \phi^{*}(B) Y_{t} \equiv \phi(B)(1-B)^{d} Y_{t}=\theta(B) Y_{t}, \quad\left\{\epsilon_{t}\right\} \sim 
N(0, \sigma^{2}), $$
where $B$ is the backshift differencing operator, $\phi(z), \theta(z)$ are polynomials of degree $p$ and $q$, respectively,
and $Z_t$ are white-noise innovations \citep{brockwell2016introduction}. The interrupted time series design, in addition to the generic ARIMA specification, also includes treatment indicators, i.e., an additional covariate with $\mathbb{I}(t > t_{\text{int}})$, where $t_{\text{int}}$ is the time of the intervention, or a dynamic treatment effect specification with $\mathbb{I}(t > t_{\text{int}}) (t-t_{\text{int}})$. We will discuss how our findings are robust to regressing against time indicators or time since intervention. 

\begin{figure}[!ht]
    \centering
\begin{subfigure}[t]{0.5\linewidth}
\includegraphics[width=\textwidth]{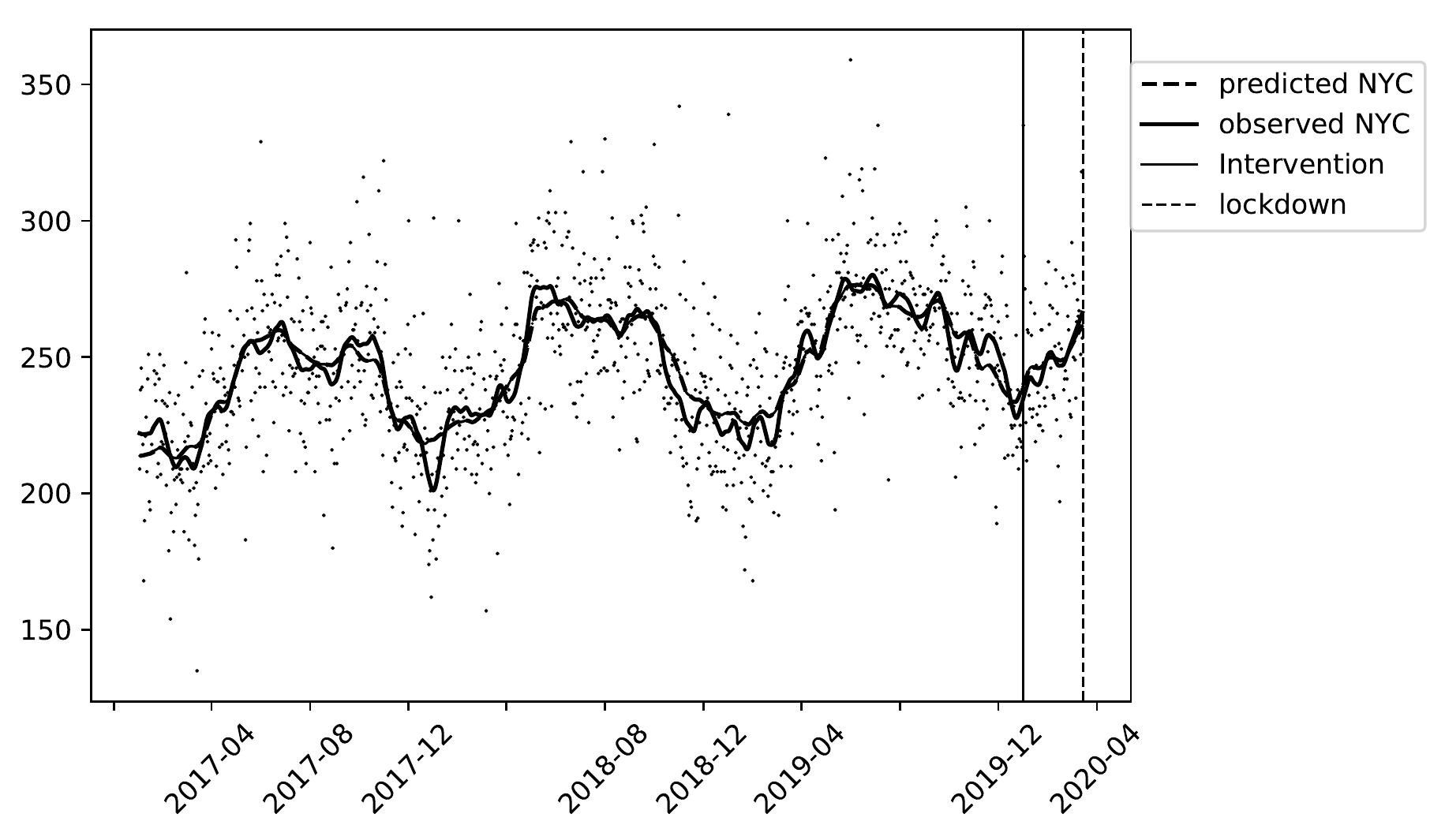}\caption{ Assault}\label{fig:ITS-ARIMA-assault}\end{subfigure}\begin{subfigure}[t]{0.5\linewidth}
\includegraphics[width=\textwidth]{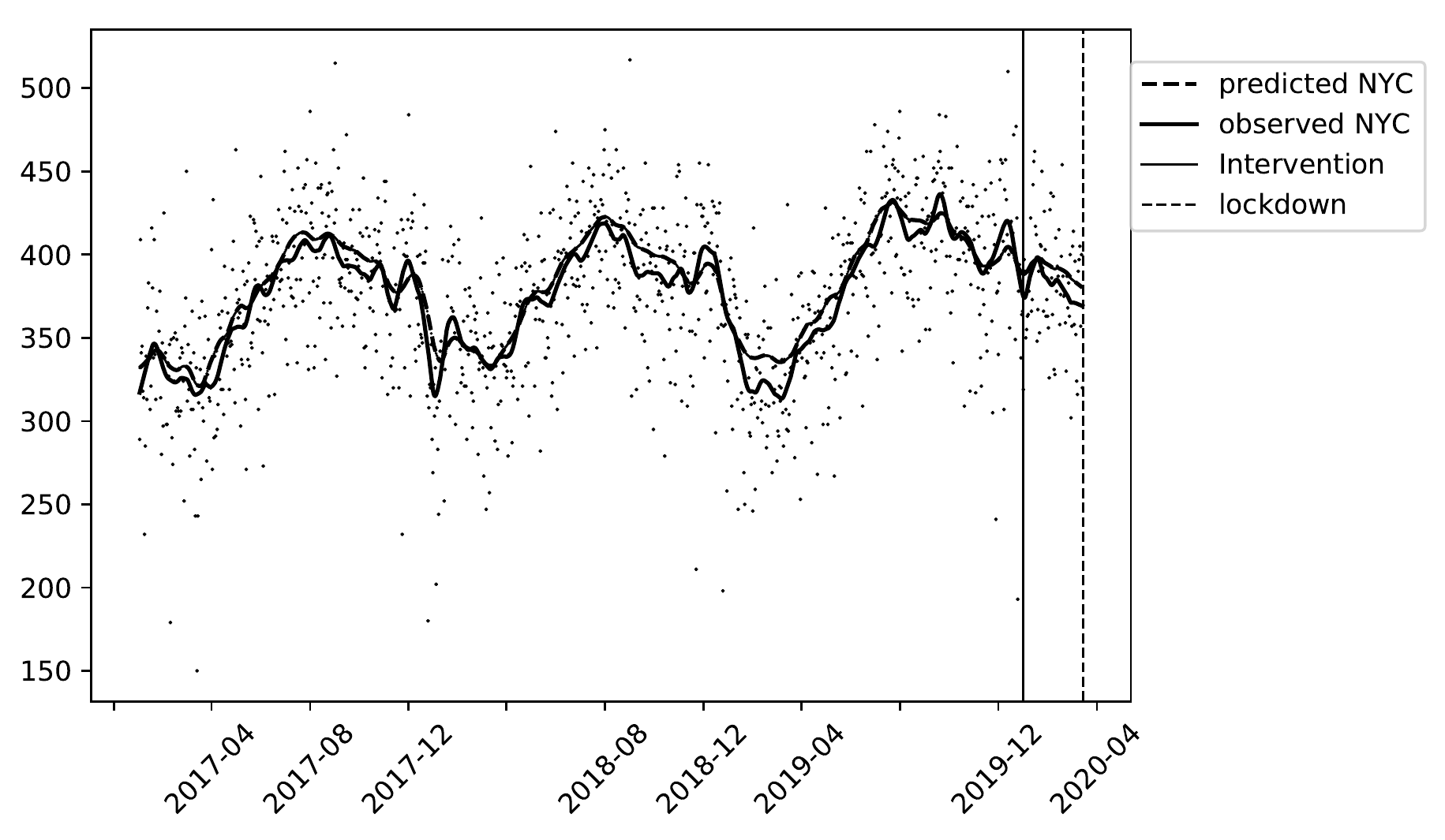}\caption{Theft}\label{fig:ITS-ARIMA-Grand-larceny}\end{subfigure}

\begin{subfigure}[t]{0.5\linewidth}
\includegraphics[width=\textwidth]{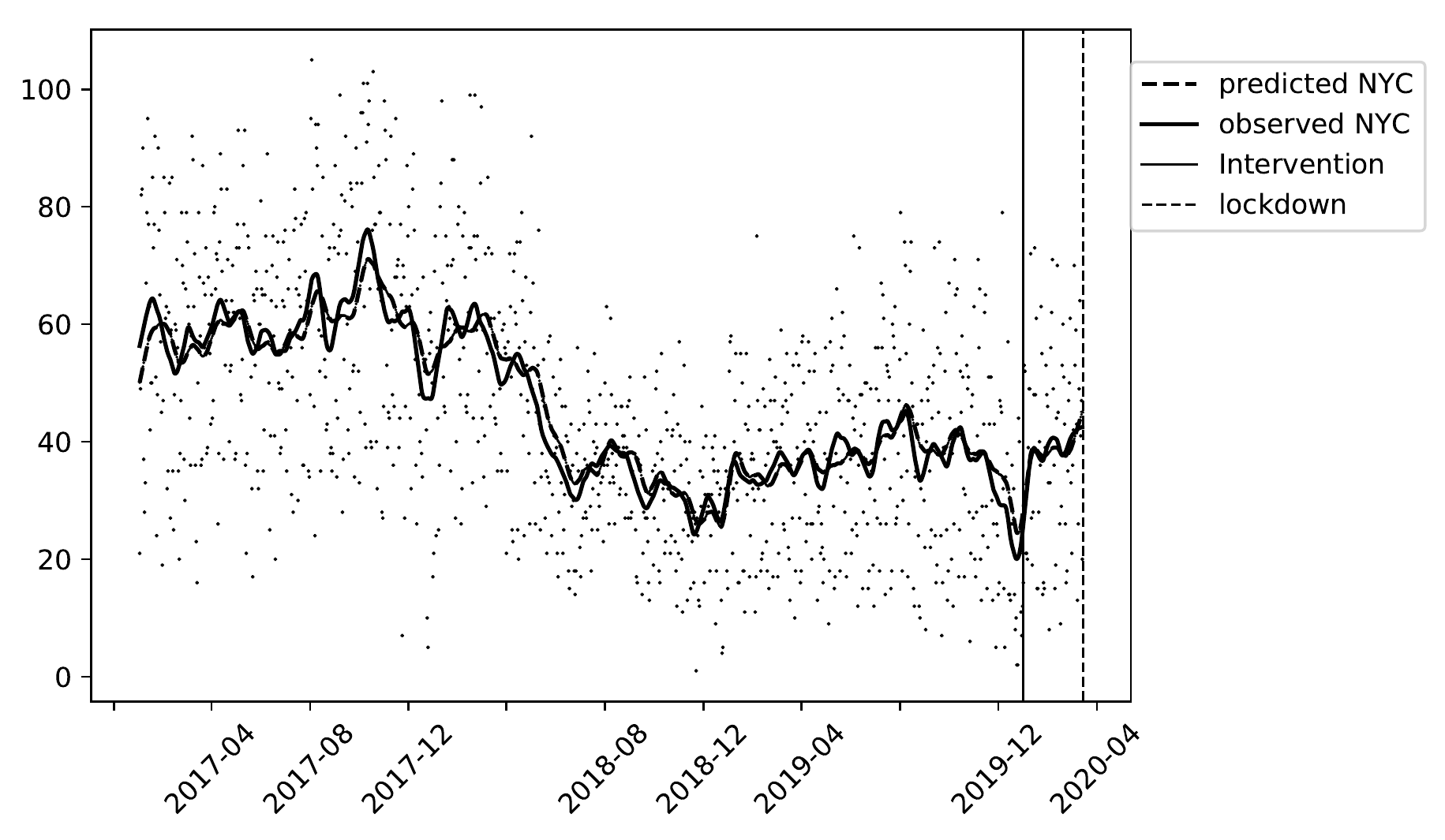}\caption{Drug}\label{fig:ITS-ARIMA-burglary}\end{subfigure}\begin{subfigure}[t]{0.5\linewidth}\includegraphics[width=\textwidth]{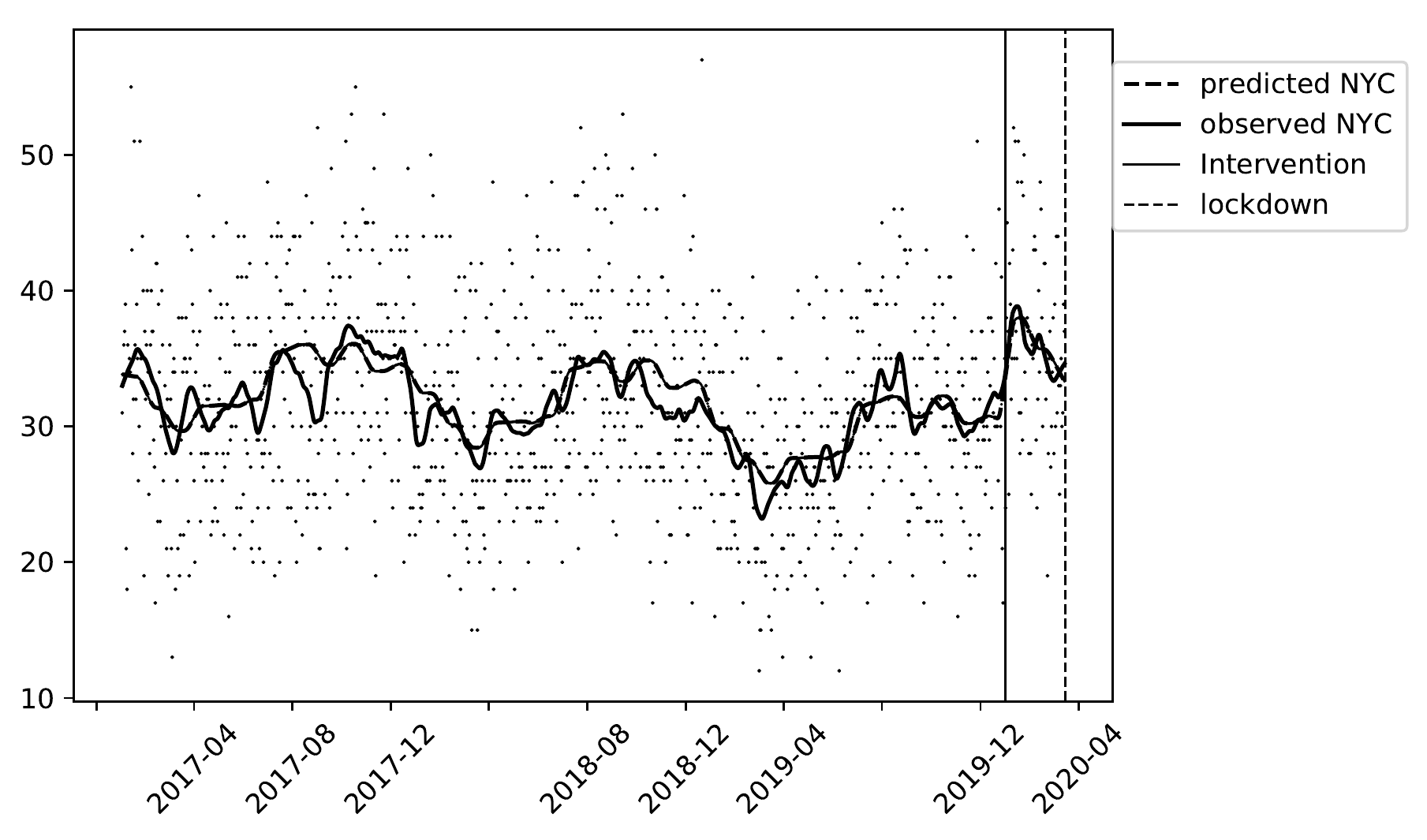}\caption{Property Burglary}\label{fig:ITS-ARIMA-Grand-larceny-mv}\end{subfigure}

\begin{subfigure}[t]{0.5\linewidth}
\includegraphics[width=\textwidth]{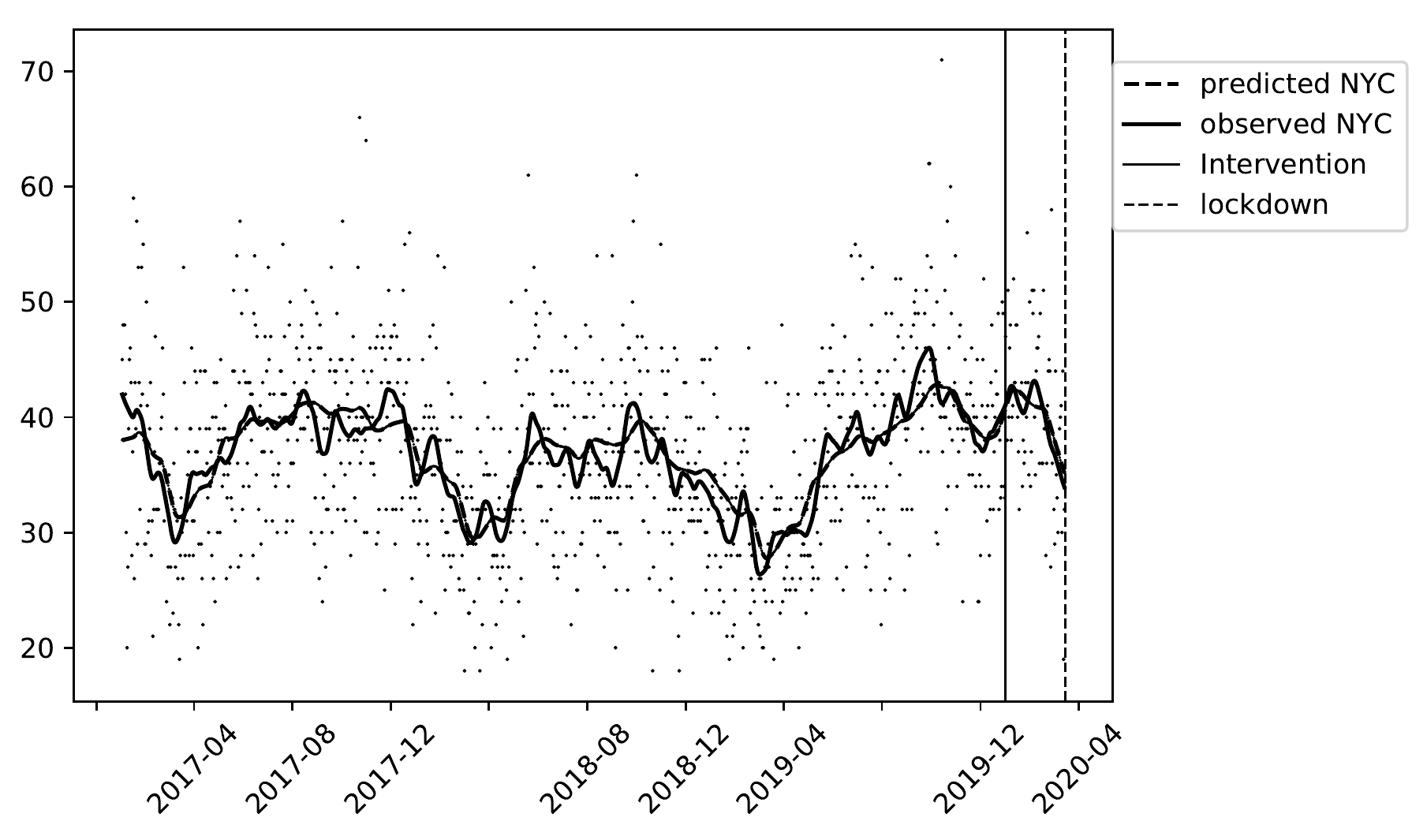}\caption{Robbery}\end{subfigure}
    \caption{ITS ARIMA plots of time series. Solid and dashed lines are loess-smoothed plots of observed and ARIMA-predicted crime counts, respectively. Scatter plot overlay of actual counts to illustrate variability. }
    \label{fig:ITS-ARIMA}
\end{figure}

We first consider an interrupted time series analysis for NYC as a within-unit case study to assess impacts of bail reform on crime rates. Because of strong temporal trends in crime rates, both seasonal within a year and within a week, and across years, interrupted time series analysis allows us to assess an increase or decrease in crime rates after the intervention in the context of decreases or increases in crime rates over time (due to seasonal fixed effects or year-over-year trends). We collect data from January 1, 2017 to March 15, 2020.\footnote{We use a longer time frame than we use for the synthetic control in the next section, where a shorter time frame allows us to compare against a larger number of different units.}

We adjust for time series structure by an ARIMA specification and with day-of-week and month fixed effects to adjust for seasonal non-stationary in each time series. We use the Hyndman-Khandakar algorithm, i.e., using unit root tests and AIC comparisons to perform model selection of the number of AR and MA terms. We include a dummy variables for ``holiday," including national holidays, New Year's Day, and Halloween. Due to data sparsity we omit the homicide and shooting time series and conduct the analysis for the same categories as we do for the synthetic control: \textit{Assault}, \textit{Theft}, \textit{Drug}, \textit{Burglary}, and \textit{Robbery}. 

In \Cref{table:its-descriptives-pre,table:its-descriptives} we include descriptive statistics of the crime time series. The \textit{Assault} and \textit{Theft} crime time series have the highest magnitude at 247 and 374 average daily cases from 2017-2020. \textit{Drug}, \textit{Robbery}, and \textit{Burglary} have an order of magnitude lower average incidence at around 45, 31, and 36 average daily cases, respectively.

We report the regression table in \Cref{table:ITS-coefficients} across models over the ARIMA models of different orders for each crime time. The treatment effect estimate is given by the regression coefficient in the ARIMA regression denoted by ``t."
The table also reports the AR/MA and drift terms found via model selection. While model selection does not suggest a time series model for the drug time series (after adjusting for day of week, month, and year fixed effects), for other crimes the specification varies. This analysis is conducted in terms of absolute number of crimes. The regression coefficient on the treatment intervention indicator (that is, only testing for a level change in crime rates) is $-66.03 \; (556.13)$, $13.78 \;(1042.05)$, $12.91 \;(39.12)$, $25.28\; (125.57)$, $-2.88\; (53.05)$, respectively, with standard errors in parentheses. Note the large standard errors. \Cref{fig:ITS-ARIMA} displays the figures to provide a visual aid of the high variability of the data. While \Cref{table:ITS-coefficients} reports as a visual aid unadjusted significance at $0.001, 0.01, 0.05$ levels, we also adjust for multiple hypothesis testing via the Holm-\v Sid\'ak correction. Without multiple adjustment, the effect on violent assault (with a $p$-value of $4.31 \times 10^{-3}$), burglary (with a $p$-value of $3.9 \times 10^{-2}$), and drug crimes (with a $p$-value of $2.41 \times 10^{-2}$) would be significant at a $0.05$ significance level.  After multiple-adjustment, only the (negative) effect on violent assault is statistically significant. However, since a decrease in crime in bail reform is not consistent with the mechanisms and taken together with the limitations discussed previously, overall the results do not suggest a change -- and in particular any positive increase -- in crime due to treatment after adjusting for multiple comparisons.

\paragraph{Homicide time series.} 
While we do not study homicide using synthetic controls due to its prohibitive sparsity (0.841 daily homicides on average), we do conduct an ITS analysis on homicide given its criminological significance.
Because of the sparse and discrete nature of the time series, we use a generalized-linear model version of ARIMA adjustment for Poisson outcomes. We report the results in \cref{table:its-homicide}. The results suggest a statistically insignificant increase in post-intervention homicides. We report results under two specifications: either level changes (treatment being the indicator variable of post-treatment intervention) or level and slope change (including additional interaction with treatment indicator and number of days post-intervention). The findings are qualitatively the same across specifications.

\begin{table}[!ht]
\vspace{-35pt}
\caption{ITS ARIMA analysis. 
}
\label{table:coefficients}
\resizebox{0.6\textwidth}{!}{\begin{tabular}{l c c c c c }
\hline
 &  Assault &  Theft & Drug &  Burglary & Robbery \\
\hline
AR(1)         & $0.47^{***}$    & $-0.23^{**}$   &               & $-0.14$         & $0.99^{***}$  \\
            & $(0.02)$        & $(0.01)$       &               & $(0.03)$        & $(0.00)$      \\
AR(2)         &                 & $0.71^{***}$   &               & $0.47^{***}$    &               \\
            &                 & $(0.01)$       &               & $(0.02)$        &               \\
AR(3)         &                 &                &               & $-0.14^{***}$   &               \\
            &                 &                &               & $(0.00)$        &               \\
MA(1)         & $-0.29^{*}$     & $0.39^{***}$   &               & $-0.65^{***}$   & $-0.88^{***}$ \\
            & $(0.02)$        & $(0.01)$       &               & $(0.03)$        & $(0.00)$      \\
MA(2)         &                 & $-0.57^{***}$  &               & $-0.80^{***}$   & $-0.08^{**}$  \\
            &                 & $(0.01)$       &               & $(0.03)$        & $(0.00)$      \\
MA(3)         &                 &                &               & $0.50^{***}$    &               \\
            &                 &                &               & $(0.02)$        &               \\
drift       &                 &                &               & $-0.29^{***}$   &               \\
            &                 &                &               & $(0.00)$        &               \\
intercept   & $315.81^{***}$  & $370.65^{***}$ & $18.85^{**}$  &                 & $46.72^{***}$ \\
            & $(552.14)$      & $(1087.62)$    & $(39.63)$     &                 & $(56.83)$     \\
t           & $-66.03^{**}$   & $13.78$        & $12.91^{*}$   & $25.28^{*}$     & $-2.88$       \\
            & $(535.13)$      & $(1042.05)$    & $(39.12)$     & $(125.57)$      & $(53.05)$     \\
Monday      & $2.15$          & $57.51^{***}$  & $6.97^{***}$  & $1.24$          & $-0.20$       \\
            & $(5.23)$        & $(10.45)$      & $(0.43)$      & $(1.26)$        & $(0.54)$      \\
Tuesday     & $2.61$          & $57.01^{***}$  & $7.36^{***}$  & $22.95^{***}$   & $-2.24^{**}$  \\
            & $(5.86)$        & $(10.94)$      & $(0.43)$      & $(1.56)$        & $(0.59)$      \\
Wednesday   & $7.60^{**}$     & $60.17^{***}$  & $8.08^{***}$  & $32.34^{***}$   & $-2.14^{**}$  \\
            & $(6.12)$        & $(11.19)$      & $(0.43)$      & $(1.59)$        & $(0.59)$      \\
Thursday    & $1.04$          & $52.16^{***}$  & $8.15^{***}$  & $30.05^{***}$   & $-2.73^{***}$ \\
            & $(6.09)$        & $(11.14)$      & $(0.43)$      & $(1.59)$        & $(0.59)$      \\
Friday      & $9.20^{***}$    & $78.08^{***}$  & $13.72^{***}$ & $29.60^{***}$   & $-1.22$       \\
            & $(5.84)$        & $(10.90)$      & $(0.43)$      & $(1.56)$        & $(0.59)$      \\
Saturday    & $-0.32$         & $35.61^{***}$  & $5.06^{***}$  & $17.22^{***}$   & $0.57$        \\
            & $(5.19)$        & $(10.38)$      & $(0.43)$      & $(1.25)$        & $(0.54)$      \\
January     & $-4.98$         & $-47.93^{***}$ & $-0.92$       & $-84.75^{***}$  & $-1.30$       \\
            & $(18.01)$       & $(44.51)$      & $(0.73)$      & $(162.74)$      & $(1.76)$      \\
February    & $-3.41$         & $-51.51^{***}$ & $-3.26^{***}$ & $-78.93^{***}$  & $-3.06^{*}$   \\
            & $(18.92)$       & $(48.60)$      & $(0.76)$      & $(141.41)$      & $(2.09)$      \\
March       & $3.62$          & $-56.99^{***}$ & $-4.97^{***}$ & $-72.34^{***}$  & $-7.00^{***}$ \\
            & $(18.55)$       & $(48.52)$      & $(0.74)$      & $(121.83)$      & $(2.18)$      \\
April       & $16.39^{***}$   & $-35.14^{***}$ & $-3.10^{***}$ & $-62.89^{***}$  & $-5.55^{***}$ \\
            & $(19.88)$       & $(50.29)$      & $(0.80)$      & $(101.30)$      & $(2.36)$      \\
May         & $37.78^{***}$   & $-13.23$       & $-3.02^{***}$ & $-55.77^{***}$  & $-0.59$       \\
            & $(19.49)$       & $(49.90)$      & $(0.78)$      & $(81.90)$       & $(2.42)$      \\
June        & $42.03^{***}$   & $7.48$         & $-2.65^{**}$  & $-51.65^{***}$  & $0.44$        \\
            & $(19.91)$       & $(50.28)$      & $(0.80)$      & $(64.37)$       & $(2.57)$      \\
July        & $35.44^{***}$   & $28.47^{***}$  & $0.93$        & $-39.75^{***}$  & $0.28$        \\
            & $(19.49)$       & $(49.23)$      & $(0.78)$      & $(50.47)$       & $(2.34)$      \\
August      & $30.79^{***}$   & $25.41^{***}$  & $1.43$        & $-29.49^{***}$  & $1.85$        \\
            & $(19.65)$       & $(49.53)$      & $(0.79)$      & $(37.10)$       & $(2.25)$      \\
September   & $36.14^{***}$   & $19.78^{**}$   & $-0.10$       & $-22.16^{***}$  & $2.64$        \\
            & $(19.87)$       & $(50.31)$      & $(0.80)$      & $(24.81)$       & $(2.14)$      \\
October     & $27.35^{***}$   & $13.09$        & $1.48$        & $-10.51^{**}$   & $3.05^{*}$    \\
            & $(19.47)$       & $(49.98)$      & $(0.78)$      & $(15.96)$       & $(1.97)$      \\
November    & $9.78^{*}$      & $-6.11$        & $-0.56$       & $-6.02^{*}$     & $0.31$        \\
            & $(19.56)$       & $(47.50)$      & $(0.80)$      & $(7.32)$        & $(1.59)$      \\
2017        & $-101.04^{***}$ & $-42.31$       & $8.57$        & $-279.88^{***}$ & $-6.67$       \\
            & $(534.30)$      & $(1044.53)$    & $(38.84)$     & $(1762.93)$     & $(55.55)$     \\
2018        & $-90.16^{***}$  & $-35.40$       & $7.34$        & $-180.77^{***}$ & $-8.87$       \\
            & $(534.12)$      & $(1044.09)$    & $(38.82)$     & $(844.79)$      & $(55.87)$     \\
2019        & $-84.78^{***}$  & $-29.36$       & $4.73$        & $-80.40^{***}$  & $-8.54$       \\
            & $(532.84)$      & $(1039.12)$    & $(38.82)$     & $(294.01)$      & $(53.58)$     \\
holiday     & $11.27^{*}$     & $-62.76^{***}$ & $-2.01$       & $-14.65^{***}$  & $3.37$        \\
            & $(32.05)$       & $(61.89)$      & $(2.24)$      & $(7.51)$        & $(3.11)$      \\
\hline
R$^2$       & 0.41            & 0.60           & 0.38          & 0.72            & 0.23          \\
Adj.\ R$^2$ & 0.39            & 0.59           & 0.37          & 0.71            & 0.21          \\
\hline
\multicolumn{6}{l}{\scriptsize{$^{***}p<0.001$, $^{**}p<0.01$, $^*p<0.05$}}
\end{tabular}
}
\end{table}

\begin{table}[!htbp]
\begin{center}
\caption{ITS: $p$-values of treatment effect regression coefficients }
\begin{tabular}{l |c c c c c }
\hline 
&Violent Assault & Theft & Burglary & Drug & Robbery\\ \hline
$p$-value &$4.31 \times 10^{-3}$ & $6.7 \times 10^{-1}$ & $3.90 \times 10^{-2}$ & $2.41 \times 10^{-2}$ & $6.93 \times 10^{-1}$\\ 
Adjusted p&$2.14 \times 10^{-2}$ & $8.91 \times 10^{-1}$ & $1.13 \times 10^{-1}$ & $9.29 \times 10^{-2}$ & $8.91 \times 10^{-1}$\\ 

\end{tabular}
\label{table:ITS-coefficients}
\end{center}
\end{table}

\begin{figure}[!ht]
    \centering
    \includegraphics[width=0.7\textwidth]{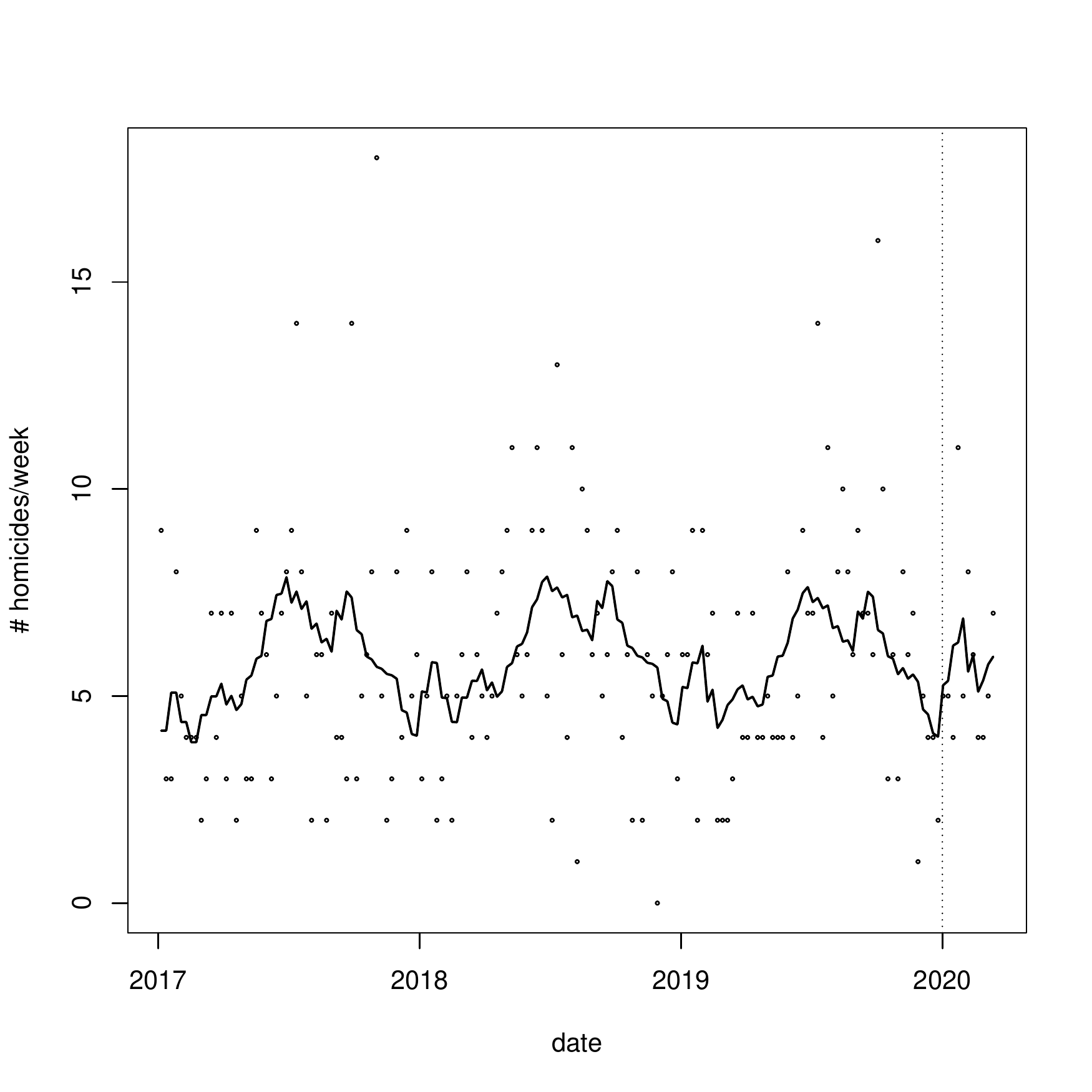}
    \caption{Homicide ITS (Poisson GLM): predictions vs. observations, weekly aggregation. Scatter plot of observed counts. Solid line of predictions. }
    \label{fig:its-homicide}
\end{figure}

\begin{table}
\begin{center}
\caption{ITS, Poisson GLM for homicide time series. $t$ is the indicator variable for post-intervention period. $t \times$ number of days is interacted with number of days post-intervention.}
\label{table:its-homicide}
\begin{tabular}{l c c }
\hline
 & Uninteracted treatment & Time-interacted treatment \\
\hline
(Intercept) & $0.95$   & $2.67^{**}$ \\
            & $(0.95)$ & $(0.95)$    \\
AR(1)    & $0.00$   & $0.00$      \\
            & $(0.05)$ & $(0.05)$    \\
AR(2)    & $0.69$   & $0.34$      \\
            & $(0.35)$ & $(0.35)$    \\
t           & $0.34$   & $0.41$      \\
            & $(1.40)$ & $(1.40)$    \\
January     & $1.23$   & $1.16$      \\
            & $(0.80)$ & $(0.65)$    \\
February    & $0.00$   & $0.00$      \\
            & $(0.65)$ & $(0.66)$    \\
March       & $0.82$   & $0.56$      \\
            & $(0.66)$ & $(0.70)$    \\
April       & $0.48$   & $0.48$      \\
            & $(0.70)$ & $(0.70)$    \\
May         & $1.24$   & $1.29$      \\
            & $(0.70)$ & $(1.00)$    \\
June        & $1.83$   & $2.06$      \\
            & $(1.00)$ & $(1.30)$    \\
July        & $1.17$   & $2.48^{*}$  \\
            & $(1.30)$ & $(0.94)$    \\
August      & $0.79$   & $1.34$      \\
            & $(0.94)$ & $(1.04)$    \\
September   & $1.73$   & $1.96$      \\
            & $(1.04)$ & $(1.02)$    \\
October     & $0.50$   & $1.31$      \\
            & $(1.02)$ & $(0.89)$    \\
November    & $0.62$   & $0.79$      \\
            & $(0.89)$ & $(1.20)$    \\
2017        & $0.00$   & $0.00$      \\
            & $(1.20)$ & $(1.27)$    \\
2018        & $0.09$   & $0.18$      \\
            & $(1.27)$ & $(1.21)$    \\
2019        & $0.01$   & $0.06$      \\
            & $(1.21)$ & $(0.95)$    \\
t $\times$ number of days        &          & $0.01$      \\
            &          & $(0.80)$    \\
\hline
R$^2$       & 0.12     & 0.11        \\
Adj.\ R$^2$ & 0.02     & -0.00       \\
Num.\ obs.  & 167   & 167      \\
\hline
\multicolumn{3}{l}{\scriptsize{$^{***}p<0.001$, $^{**}p<0.01$, $^*p<0.05$}}
\end{tabular}
\end{center}
\end{table}

\clearpage
\section{Robustness checks}\label{sec:apx-robustnesschecks}

\subsection{Uncertainty quantification and robustness checks}\label{apx-robustnesschecks}

\paragraph{Placebo checks in time} 

Another placebo check suggested in \citet{abadie2011synth} considers changing the intervention time and verifying estimation of a null effect. We change the intervention time to an earlier date, and re-run the analysis considering the period from the new intervention time to the old intervention time as the placebo post-intervention period. We expect to verify that the synthetic control analysis suggests a null effect for these redefined post-intervention times, which are in fact all previous to the actual implementation time.

In \Cref{tab:sc-ridge-placebos-time}, we present the in-time placebo check. We modify the intervention period to start earlier by a year, six months, and three months: one of 01/01/19, 06/01/19, or 09/01/19. We then define the time period between the new intervention and the original intervention timepoint, 01/01/2020, as the placebo ``post-intervention" time period, and compute similar placebo $p$-values. Since we know these timepoints do not correspond to interventions, they provide estimates of the synthetic control ATE under a known null model of no effect. The corresponding ``null ATEs" are of similar magnitude to the observed ATEs under the synthetic control suggesting robust null conclusions. 

The placebo $p$-value quantifies the $p$-values under the null model. 
We also report the average placebo RMSE to verify that these placebo checks achieve similar overall predictive fit, of similar orders of magnitude to the original synthetic control analysis. 
Note that the ATE test statistic is one sided, so the least extreme $p$-value is $0.5$, while for the RMSE test statistic, even larger $p$-values indicate that the NYC post-treatment RMSE is even less ``extremal" within the placebo distribution.

\paragraph{Robustness checks: early roll-in of intervention to assess anticipation effects} 
In \Cref{tab:sc-ridge-placebos-earlyrollin} we present the robustness checks corresponding to early roll-in of intervention.

In the specific context of bail reform, descriptive statistics on the frequency of ROR (released on one's own recognizance) cases suggest that judges were incorporating the bail reform requirements for release before the official January 1 implementation date, that is, there was anticipation of the intervention \citep{cci_summary}. Therefore, to test robustness of our conclusions to the anticipation effect, we also re-run the synthetic control analysis where we \textit{move-up} the start date of the intervention itself to November, October, or September. 

For example, in a hypothetical case where treatment effects on crime were only short-term, perhaps the early phase-in of bail reform requirements by judges would result in a statistically significant treatment effect on crime, but only measuring post-January data would hide the effect. On the other hand, early phase-in could have also increased the time at risk for some of those who were released prior to the intervention roll-in, inflating our observed effect. To some extent, using synthetic control analysis rather than a discontinuity-local analysis (such as ITS) mitigates potential impacts of the combination of dynamic treatment effects and anticipation, since synthetic control fits the few weights to minimize error over a larger time horizon, the entire pre-treatment period rather than locally near the boundary.

This ``early roll-in" robustness check differs from the previous ``in-time" placebo check because we increase the post-intervention time-frame by moving the intervention time earlier, while the ``in-time" placebos re-run synthetic control entirely with pre-intervention data. Therefore, rather than necessarily expecting null effects as in the ``in-time" checks, instead robustness here would be exhibited by seeing effects that are in general agreement with the main analysis, whether null or not.

\begin{table}[ht!]
    \centering
        \caption{Placebos in time; Ridge }
    \begin{tabular}{ll|llll}
    Category & Start date & ATE (/1000) &  p (ATE) & p (RMSE) & Avg. Placebo RMSE\\ \hline
 Assault&01/01/19&$-0.0195$ & $0.86$ & $0.81$ & $2.24 \times 10^{-5}$\\
&03/01/19&$-0.0084$ & $0.64$ & $1.0$ & $2.31 \times 10^{-5}$\\
&06/01/19&$-0.0056$ & $0.59$ & $1.0$ & $2.47 \times 10^{-5}$\\
 Theft&01/01/19&$-0.0315$ & $0.74$ & $1.0$ & $3.74 \times 10^{-5}$\\
&03/01/19&$-0.0074$ & $0.74$ & $1.0$ & $4.11 \times 10^{-5}$\\
&06/01/19&$0.0216$ & $0.45$ & $1.0$ & $4.12 \times 10^{-5}$\\
 Burglary&01/01/19&$0.0047$ & $0.26$ & $0.59$ & $1.78 \times 10^{-5}$\\
&03/01/19&$0.0035$ & $0.39$ & $0.55$ & $1.89 \times 10^{-5}$\\
&06/01/19&$0.0033$ & $0.35$ & $0.86$ & $1.91 \times 10^{-5}$\\
 Robbery&01/01/19&$0.0037$ & $0.13$ & $0.27$ & $1.19 \times 10^{-5}$\\
&03/01/19&$0.0039$ & $0.22$ & $0.18$ & $1.21 \times 10^{-5}$\\
&06/01/19&$0.0048$ & $0.22$ & $0.41$ & $1.24 \times 10^{-5}$\\
Drug&01/01/19&$-0.0005$ & $0.35$ & $1.0$ & $2.06 \times 10^{-5}$\\
&03/01/19&$0.0012$ & $0.35$ & $1.0$ & $2.14 \times 10^{-5}$\\
&06/01/19&$0.003$ & $0.24$ & $1.0$ & $2.20 \times 10^{-5}$\\
    \end{tabular}
    \label{tab:sc-ridge-placebos-time}
\end{table}

\clearpage

\begin{table}[]
	\centering
	\caption{Placebo, Ridge (early roll-in of intervention start) }
	\begin{tabular}{ll|llll}
	    \hline
    \hline
Category & Early Roll-in Start date & ATE (/ 1000) & p (ATE) & p (RMSE) & Avg. Placebo RMSE\\ \hline
 Assault&09/01/19&$-0.0002$ & $0.52$ & $1.0$ & $2.45 \times 10^{-5}$\\
&10/01/19&$-0.0006$ & $0.52$ & $1.0$ & $2.47 \times 10^{-5}$\\
&11/01/19&$0.0045$ & $0.52$ & $1.0$ & $2.48 \times 10^{-5}$\\
 Theft&09/01/19&$0.0354$ & $0.32$ & $0.95$ & $4.30 \times 10^{-5}$\\
&10/01/19&$0.0203$ & $0.36$ & $1.0$ & $4.39 \times 10^{-5}$\\
&11/01/19&$0.0186$ & $0.27$ & $1.0$ & $4.47 \times 10^{-5}$\\
 Burglary&09/01/19&$0.0005$ & $0.39$ & $0.86$ & $1.96 \times 10^{-5}$\\
&10/01/19&$0.0014$ & $0.39$ & $0.77$ & $1.96 \times 10^{-5}$\\
&11/01/19&$0.0023$ & $0.43$ & $0.45$ & $2.00 \times 10^{-5}$\\
 Robbery&09/01/19&$0.0066$ & $0.13$ & $0.09$ & $1.28 \times 10^{-5}$\\
&10/01/19&$0.0062$ & $0.09$ & $0.18$ & $1.29 \times 10^{-5}$\\
&11/01/19&$0.006$ & $0.04$ & $0.32$ & $1.29 \times 10^{-5}$\\
Drug&09/01/19&$-0.0008$ & $0.44$ & $0.73$ & $1.98 \times 10^{-5}$\\
&10/01/19&$-0.002$ & $0.44$ & $0.67$ & $1.96 \times 10^{-5}$\\
&11/01/19&$-0.0049$ & $0.56$ & $0.4$ & $1.96 \times 10^{-5}$\\
\end{tabular}
    \label{tab:sc-ridge-placebos-earlyrollin}
\end{table}

\clearpage

\clearpage

\section{Description of dataset}\label{sec:apx-datasetdescription}
\singlespacing

\subsection{Comparison to UCR Part I index crime aggregates}\label{sec:apx-datasetdescription-subUCR}

We discuss how our crime categorization differs from UCR Part I index crimes. For the most part, descriptions of the broad crime categories align with UCR, although there are differences. The largest discrepancies are in the theft and assault categories. Assault differs the most from UCR designations and includes non-felony assault (simple assault) since some jurisdictions do not differentiate reports between felony/non-felony assaults. Our ``theft" designation includes both felony and misdemeanor crimes. In \Cref{tab:nycdatamapping-violent} we include a full description of the crime descriptors reported in NYC (offense description, law description, and police department descriptions) describing which crimes are aggregated to our final categories. 

While many cities have switched to NIBRS reporting, where multiple crime types can be reported from a single incident, some cities may report based on Uniform Crime Reporting (UCR) Traditional Summary reporting, where only the most severe crime type is reported per incident. Our dataset is simply an aggregation of each city’s crime regardless of reporting schemes, so some cities may appear to have more crime than others based on this difference in reporting. That is, we did not manually align reporting across cities according to the hierarchy rule. Moreover, all cities do not report crime types at the same granularity of description (assault vs. aggravated/simple/unclassified assault). 

To give a sense of possible discrepancies in frequencies due to reporting with or without the hierarchy, note that an analysis of the NIBRS vs. UCR Summary Reporting system submissions from 1991 to 2011 shows that changing to NIBRS reporting, which allows multiple offenses, results in no effect on rapes, 0.5\% increase in robbery, 0.6\% increase in aggravated assault, 0.8\% increase in burglary, 3.1\% increase in larceny, and 3.8\% increase in motor vehicle theft \citep{rantala2000effects}. Only 9.2\% of reports contained more than one offense per incident. Because our analysis does not directly require comparing magnitudes of crime across different cities, we expect this does not affect our results. 

\subsection{Further details on construction}
\begin{figure}
    \centering
\includegraphics[width=\textwidth]{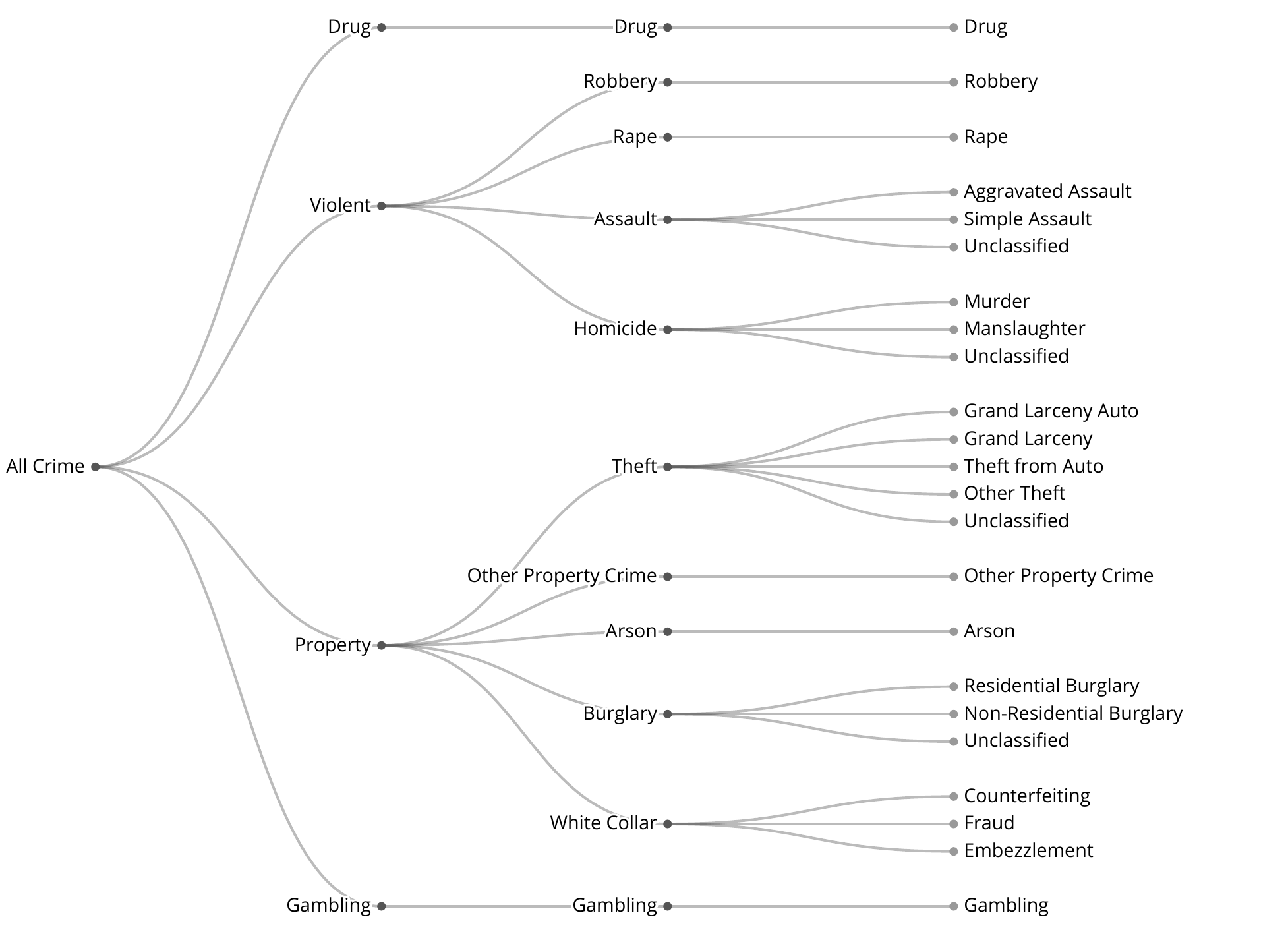}
    \caption{Overview of crime reporting hierarchy}
    \label{fig:hierarchy}
\end{figure}

\textit{Level 1} of the hierarchy describes broad categorizations of crime types: violent, drug, property crime, gambling (and other). \textit{Level 2} includes further categorization: drug, robbery, rape, assault, homicide, theft, other property crime, arson, burglary, white collar, and gambling. \textit{Level 3}, which we do not use, describes further subtypes, for example aggravated vs. simple assault or residual vs. non-residential burglary. We conduct our analysis at the second level of categorization. Of these crime types, the drug, robbery, assault, theft, burglary crime types had sufficient event frequency to conduct the analysis. 
\Cref{fig:hierarchy} describes the data reporting hierarchy that was used to merge data across reporting agencies. The figure includes some crimes that were not used for the final synthetic control analysis. 

The second level of the hierarchy, at which level we conducted our analysis is: 
\begin{itemize}
    \item Homicide: Murder and manslaughter (does not include justifiable homicide)
    \item Rape: Forcible or aggravated penetration (does not include statutory rape)
    \item Robbery: All taking of property through force or threat of force
    \item Assault: Includes aggravated and simple assaults when possible
    \item Burglary: Unlawful entry into structure (residential and commercial) to commit theft without use of force
    \item Theft: Unlawful taking of property (includes motor vehicle theft)
    \item Other property crime: Includes a broad set of property crimes such as arson, stolen property offenses, damage to property, trespass, and other miscellaneous property crimes. Excludes animal crimes, drug, theft, burglary, and white collar crimes.
    \item Drug: All drug-related abuses including cultivation, distribution, sale, purchase, use, possession, transportation, or importation of any controlled drug or narcotic substance.
    \item White collar: Includes crimes of deceit or intentional misrepresentation, such as counterfeiting, fraud, and embezzlement (includes offenses such as check fraud, confidence game, and credit card fraud)
    \item Gambling: All unlawful betting or wagering, tampering, or operation of game of chance (includes equipment violations)
    \item Arson: Any willful or malicious burning or attempt to burn property

\end{itemize}

We aggregated the incident descriptions under these categories. Some, but not all cities, reported UCR codes which were used when available. Austin, Chicago, Cincinnati, Dallas, Los Angeles, Philadelphia reported UCR codes; but not all at the same granularity (with some reporting coarse codes at the hundreds level). Houston's schema changed twice during the study period; crime counts were discontinuous across schema changes under the aggregations and so we excluded Houston from the analysis. 

These definitions are modified from other crime frameworks, such as a previous Bureau of Justice Statistics analysis\footnote{\url{https://www.bjs.gov/recidivism/}}, FBI offense definitions \footnote{\url{https://ucr.fbi.gov/crime-in-the-u.s/2019/crime-in-the-u.s.-2019/topic-pages/offense-definitions}}, and definitions from City Crime Stats \footnote{\url{https://papers.ssrn.com/sol3/papers.cfm?abstract_id=3674032}}.

\clearpage

\begin{table}[]
\caption{Data sources} 
\label{tab:datasources} 
\begin{tabular}{ll}
City           & Data Source                                                                                    \\
\hline
Atlanta        & http://opendata.atlantapd.org/Crimedata/Default.aspx                                              \\
Austin         & https://data.austintexas.gov/Public-Safety/Crime-Reports/fdj4-gpfu                                \\
Baltimore      & https://www.baltimorepolice.org/crime-stats/open-data                                             \\
Boston         & https://data.boston.gov/dataset/crime-incident-reports-august-2015-to-date-source-new-system      \\
Buffalo        & https://data.buffalony.gov/Public-Safety/Crime-Incidents/d6g9-xbgu                                \\
Chicago        & https://data.cityofchicago.org/Public-Safety/Crimes-2001-to-Present/ijzp-q8t2                     \\
Cincinatti     & https://data.cincinnati-oh.gov/Safety/PDI-Police-Data-Initiative-Crime-Incidents/k59e-2pvf        \\
Dallas         & https://www.dallasopendata.com/Public-Safety/Police-Incidents/qv6i-rri7                           \\
Denver         & https://www.denvergov.org/opendata/dataset/city-and-county-of-denver-crime                        \\
Detroit        & https://data.detroitmi.gov/datasets/rms-crime-incidents                                           \\
Fort Worth     & https://data.fortworthtexas.gov/Public-Safety/Crime-Data/k6ic-7kp7                                \\
Houston        & https://www.houstontx.gov/police/cs/index-2.htm                                                   \\
Kansas City    & https://data.kcmo.org/Crime/KCPD-Crime-Data-2020/vsgj-uufz                                        \\
Los Angeles    & https://data.lacity.org/A-Safe-City/Crime-Data-from-2020-to-Present/2nrs-mtv8                     \\
Louisville     & https://data.louisvilleky.gov/dataset/crime-reports                                               \\
Milwaukee      & https://data.milwaukee.gov/dataset/wibr                                                           \\
Nashville      & https://data.nashville.gov/Police/Metro-Nashville-Police-Department-Incidents/2u6v-ujjs           \\
New York City  & https://data.cityofnewyork.us/Public-Safety/NYPD-Complaint-Data-Current-Year-To-Date-/5uac-w243   \\
Philadelphia   & https://www.opendataphilly.org/dataset/crime-incidents                                            \\
Phoenix        & https://www.phoenixopendata.com/dataset/crime-data/resource/0ce3411a-2fc6-4302-a33f-167f68608a20  \\
Portland       & https://www.portlandoregon.gov/police/71978                                                       \\
Raleigh        & https://data-ral.opendata.arcgis.com/datasets/ral::raleigh-police-incidents-nibrs/about           \\
Sacramento     & https://data.cityofsacramento.org/datasets/0026878c24454e16b169b3fb26130751\_0/explore            \\
San Francisco  & https://data.sfgov.org/Public-Safety/Police-Department-Incident-Reports-2018-to-Present/wg3w-h783 \\
Seattle        & https://data.seattle.gov/Public-Safety/SPD-Crime-Data-2008-Present/tazs-3rd5                      \\
Virginia Beach & https://data.vbgov.com/dataset/police-incident-reports                                            \\
Washington     & https://opendata.dc.gov/datasets/crime-incidents-in-2018                                         
\end{tabular}
\end{table}

\clearpage

\clearpage
\label{tab:nycdatamapping-violent}

\begin{longtable}{l|llp{18em}}
\caption{Data mapping comparison: NYC's individual incident descriptions and our Level 2 categorization. For Police Department Description, we only include the prefix (first word) from the first description, and include the suffixes also included under the same Offense Description. Each row corresponds to a distinct cirme prefix in the Police Department Description. All terms are replicated here verbatim from the data file. }
\\

Standardized & NYC                              &           &                                                                       \\
Level 2      & Offense Description              & Law desc. & Police Department Description(s)      \\\hline
Homicide     & \makecell[l]{Murder \& Non-negl\\ Manslaughter} & Felony    &             \\
Homicide     & \makecell[l]{Homicide, \\negligent, vehicle}       & Felony    & {Homicide Negligent, Vehicle}                                                     \\
Homicide     & \makecell[l]{Homicide, \\negligent,unclassifie}   & Felony    & {Homicide,  negligent, unclassifie}                                                      \\
Rape         & Rape                             & Felony    & Rape 1                                                                         \\
Rape         & Sex Crimes                       & Felony    & Sodomy 1,2,3                                                                                                                  \\
Robbery      & Robbery                          & Felony    & 
{Robbery, Chain Store, Payroll, Atm Location, Bank, Bar/restaurant, Begin As Shoplifting, Bicycle, Bodega/convenience Store, Car Jacking, Check Cashing Business, Clothing, Commercial Unclassified, Delivery Person, Doctor/dentist Office, Dwelling, Gas Station, Hijacking, Home Invasion, Licensed For Hire Vehicle, Licensed Medallion Cab, Liquor Store, Neckchain/jewelry, Of Truck Driver, On Bus/ Or Bus Driver, Open Area Unclassified, Personal Electronic Device, Pharmacy, Pocketbook/carried Bag, Public Place Inside, Residential Common Area, Unlicensed For Hire Vehicle} \\
Assault      & Felony Assault                   & Felony    & Assault 2,1,unclassified                                                         \\
Assault      & Miscellaneous Penal Law          & Felony    & Aggravated Harassment 1                                                                                                        \\
             & Harrassment 2                    & Violation &                                                                        \\
             & Harrassment 2                    & Violation &                             \\
Burglary     & Burglary                       & Felony      & 
\makecell[l]{Burglary, Truck Unknown Time; truck Day; \\truck Night; commercial,day; commercial,night;
\\commercial,unknown Ti; unclassified,day}                              \\
Burglary     & Burglary                       & Felony      & 
\makecell[l]{Burglary,unclassified,night;\\ unclassified,unknown; unknown Time}                  \\
Theft        & \makecell[l]{Grand Larceny \\Of Motor Vehicle} & Felony      & 
Larceny, Grand Of Auto - Attem; Moped; Auto; Motorcycle; Truck                                     \\
Theft        & Grand Larceny                  & Felony      & 
{Larceny,grand By Acquiring Los; By Acquiring Lost Credit Card; By Bank Acct Compromise-atm Transaction; By Bank Acct Compromise-reproduced Check; By Bank Acct Compromise-teller; By Bank Acct Compromise-unauthorized Purchase; By Bank Acct Compromise-unclassified; By Credit Card Acct Compromise-existing Acct; By Dishonest Emp; By Extortion; By False Promise-in Person Contact; By False Promise-not In Person Contact; By Identity Theft-unclassified; By Open Bank Acct; By Open Credit Card (new Acct); By Open/compromise Cell Phone Acct; By Theft Of Credit Card; From Building  (non-residence) Unattended; From Eatery, Unattended; From Night Club, Unattended; From Open Areas, Unattended; From Person, Bag Open/dip; From Person,lush Worker(sleeping/uncon Victim); From Person,personal Electronic Device(snatch); From Person,pick; From Person,purs; From Person,uncl; From Pier, Unattended; From Residence, Unattended; From Residence/building,unattended, Package Theft Inside; From Residence/building,unattended, Package Theft Outside; From Retail Store, Unattended; From Store-shopl; Of Bicycle; Of Boat; Of Vehicular/motorcycle Accessories; Person,neck Chai} \\
Theft        & Grand Larceny                  & Felony      & Larceny,grand From Truck, Unattended; From Vehicle/motorcycle; From Boat, Unattended                                                       \\
Theft        & Petit Larceny                  & Misdemeanor & Larceny,petit From Auto; From Boat; From Truck                                           \\
Theft        & Petit Larceny                  & Misdemeanor & Larceny,petit By Acquiring Los; By Check Use; By Credit Card U; By Dishonest Emp; By False Promise; From Building,un; From Building,unattended, Package Theft Inside; From Building,unattended, Package Theft Outside; From Coin Machin; From Open Areas,; From Pier; From Store-shopl; Of Bicycle; Of Boat; Of License Plate; Of Vehicle Acces; Of Animal; check From Mailb; Of Auto - Attem; Of Moped; Of Auto; Of Motorcycle; Of Truck                                                \\
Theft        & \makecell[l]{Other Offenses \\Related To Theft} & Misdemeanor & Theft Of Services, Unclassifie; related Offenses,unclass                                                                          \\
Drug         & Dangerous Drugs                & Felony      & Controlled Substance, Intent T; Controlled Substance, Possessi; Controlled Substance, Sale 4; Controlled Substance, Sale 5; Controlled Substance,intent To; Controlled Substance,possess.; Controlled Substance,possess.-; Controlled Substance,sale 1; Controlled Substance,sale 2; Controlled Substance,sale 3; Drug Paraphernalia, Possesse; Drug, Injection Of; Marijuana, Possession 1, 2 \& 3; Marijuana, Sale 1, 2 \& 3; Sale School Grounds; Sale School Grounds 4; Sales Of Prescription                                                                                                       \\
Drug         & Dangerous Drugs                & Misdemeanor & Controlled Substance, Possessi; Drug Paraphernalia, Possesse; Marijuana, Possession 4 \& 5; Marijuana, Sale 4 \& 5; Poss Meth Manufact Material; Possession Hypodermic Instrume

\end{longtable}
 Note: The first column describes the level-2 offense. The right three columns describe incident data categorizations that comprise the level 2 mapping in our hierarchy. We omit the top-line offense description that re-appears in the more detailed police department description. These descriptions are verbatim from the raw data.

\end{document}